\documentclass{emulateapj}

\usepackage{graphics,epsfig}

\begin{document}

\title{Morphology and dynamics of solar prominences from 3D MHD simulations} 
\author{J. Terradas$^1$, R. Soler$^1$, M., Luna$^{2, 3}$, R. Oliver$^1$, J. L.
Ballester$^1$} \affil{$^1$Departament de F\'\i sica, Universitat de les Illes
Balears, E-07122 Palma de Mallorca, Spain} \affil{$^2$ Instituto de Astrof\'\i
sica de Canarias, E-38200 La Laguna, Tenerife, Spain} \affil{$^3$Departamento de
Astrof\'\i sica, Universidad de La Laguna, E-38206 La  Laguna, Tenerife, Spain} 
\email{jaume.terradas@uib.es}

\begin{abstract}

In this paper we present a numerical study of the time evolution of solar
prominences embedded in sheared magnetic arcades. The prominence is represented
by a density enhancement in a background stratified atmosphere and is connected
to the photosphere through the magnetic field. By solving the ideal
magnetohydrodynamic (MHD)  equations in three dimensions we study the dynamics
for a range of parameters representative of real prominences. Depending on the
parameters considered, we find prominences that are suspended above the
photosphere, i.e., detached prominences,  but also configurations resembling
curtain or hedgerow prominences whose material continuously connects  to the 
photosphere. The plasma$-\beta$ is an important parameter that determines the
shape of the structure. In many cases magnetic Rayleigh-Taylor (MRT)
instabilities and oscillatory phenomena develop.  Fingers and plumes are
generated, affecting the whole prominence body and producing vertical structures
in an essentially horizontal magnetic field. However, magnetic shear is able to
reduce or even to suppress this instability. 

\end{abstract}


\maketitle

\section{Introduction} 

Quiescent prominences are large structures of cool and dense plasma suspended in
quiet regions of the solar corona. These structures can have lifetimes of weeks
although they have a highly dynamic nature. It is generally believed that the
mass in a quiescent prominence is supported against gravity by the magnetic
Lorentz force. Different models of the magnetic structure of prominences have
been proposed in the past \citep[see the review of][]{mackayetal10}, and
magnetic dips, i.e., sites where the magnetic field lines are locally horizontal
and curved in the upward direction, are thought to play a relevant role.
Examples of such configurations are the \citet{kippschl57} model, hereafter
referred as KS, or the \citet{hoodanzer90} model where dips are self
consistently created by the weight of the heavy prominence. An alternative is
that dips already exist before the dense material is deposited in them. In this
context, there are models suggesting that prominences are supported by flux
ropes that are essentially horizontal and lie above the polarity inversion line,
\citep[see][]{sakurai1976,low81, rustkumar1994,rustkumar1996, lowzhang04}.
Another possibility is that instead of a flux rope it is a sheared arcade that
is responsible for the dips, c.f., 
\citet[][]{antiochosetal1994,aulanier02,devoreetal2005,karpenetal2005,lunaetal12a}.
However, the interplay between magnetic field and the dense prominence is not
addressed in these works. In the present paper we investigate in a consistent
way a magnetic shear arcade model together with a  dense prominence, leaving the
analysis of flux rope models for future studies.

The {\em Hinode} satellite has provided  unprecedented high-resolution images of quiescent
prominence allowing a detailed study of the dynamics of these structures. Prominences show
changes in morphology, irregular motions on different spatial and temporal scales,
downflows, upflows, vortexes, raising bubbles, plumes, etc. 
\citet{bergeretal08,bergeretal10} reported dark upflows that formed at the base of some
quiescent prominences. \citet{bergeretal10} proposed that these observed upflows were
caused by the Rayleigh-Taylor instability \citep[see][]{rayleigh1883,taylor50}.
\cite{ryuetal10} showed how many fundamental plasma instabilities can be linked to
dynamical processes occurring in prominences. In particular they showed that the
development of regular series of plumes and spikes taking place at the interface between
the prominence and corona are most likely related to the development of magnetic
Rayleigh-Taylor (MRT) instabilities. \citet{hillieretal11} described how upflows can be
the consequence of the MRT instability acting on the boundary between the KS prominence
model and a tube inserted in the structure. Later, \citet{hillieretal12} performed a
detailed description of the dynamics of the system using the KS model for a wide rage of
model parameters. \citet{hillieretal12a} have extended the previous works to include
interchange reconnection in the models. \citet{dudiketal2012} have
proposed that the existence of prominence bubbles is connected to a pair of magnetic null
points associated with the prominence feet, and that a separator-reconnection scenario,
which is very different from the Rayleigh-Taylor hypothesis, may naturally explain the
bubble and plume formation. The role of neutrals on the MRT instability was investigated
(in the linear regime) by \citet{diazetal2012} while   \citet{khomenkoetal2014} have
considered the nonlinear regime and have found that the configuration is always unstable
on small scales. The models presented by these authors are based on local models of
prominences and the effect of the corona is not included. Moreover, the magnetic field
does not connect to the photosphere which is an important boundary condition for
prominence support. In this paper these points are properly addressed. 

Another example of dynamical phenomena reported in quiescent prominences are
oscillations. These oscillations are either of global nature, producing motions of the
whole prominence \citep[see][]{tripatetal09, lizhang2012,lunaetal2014,shenetal2014}, or
they are local. In this last case, the small amplitude periodic motions are mostly
detected in Doppler shifts of spectral lines, see for example
\citet{tsubaki1988,oliver1999,oliball2002},  and the reviews of \citet{mackayetal10} and
\citet{arreguietal12}. The theoretical understanding of the oscillations is based on the
hypothesis that they correspond to magnetohydrodynamic (MHD) waves. An intense modeling
of prominence oscillations has been done in the last years and it has been mostly based
on the determination of the normal modes of  different prominence models. These
equilibrium models are rather simple and the configuration is represented by slabs or
cylindrical tubes. It is important to remark that the MHD perturbations on these models
are imposed to satisfy line-tying conditions at the photosphere. This is a crucial
boundary condition that has a strong effect on the eigenmodes of the prominence and has
important consequences regarding the stability of the structure. The importance of
line-tying conditions on prominence oscillations is well known, and  a self-consistent
way of investigating the dynamics of prominences must involve the effect of the
underlying photosphere \citep[see][]{schutgens1997,vanderoordetal1998}.

In \citet{terradasetal2013} a two-dimensional prominence model embedded in a
coronal quadrupolar arcade was constructed to mimic a normal polarity
prominence. In that work, dense material representing the prominence was
injected in an initially stable magnetic structure increasing the  gravitational
energy. From the analysis of the evolution of the system it was found that in
some cases the system  relaxed  towards a situation that was close to a
stationary state. The obtained models using this method represent cool
prominences supported against gravity by magnetic dips \citep[see also the work
of][who considered a two-dimensional flux rope model instead of an arcade
configuration]{hilliervan13}. The formation process \citep[see the works
of][]{xiaetal11, xiaetal12,keppensetal2014} was not addressed in this model  and
the main aim was to study MHD waves on the generated models.

The main motivation in the present paper is to extend the study of the
two-dimensional model of \citet{terradasetal2013} to the three-dimensional case.
Again our interest is on global models of prominences rather than in the details
of the internal structure. Line-tying conditions are applied at the photosphere
as they are crucial to have magnetic support and they strongly affect the
dynamics of the system. Since the model is three-dimensional it allows us to
analyze perturbations along the longitudinal axis of the prominence neglected in
the 2D case. This changes significantly the physics of the system because MRT
unstable modes are easily excited. The development of the instabilities together
with the oscillatory behavior of the structure generate complex motions in the
prominence body. One of the main aims of this work is to investigate the
dynamics but also to understand how the morphology  of the prominence depends on
the different parameters of the model.

\section{Initial configuration and setup}

The basic initial configuration is an isothermal stratified atmosphere as in
\citet{terradasetal2013}. The same parameters have been used in the present work. The main
difference is in the magnetic field configuration, which now has a component in the
$y-$direction that introduces shear in the system.  Again the force-free magnetic field is
based on arcade solutions. The magnetic field has to satisfy
\begin{eqnarray}\label{helm} 
\left(\nabla^2 +\alpha^2\right){\bf B}=0,
\end{eqnarray}
where it has been assumed that $\nabla\times{\bf B}=\alpha {\bf B}$ and $\alpha$ uniform.
 
An arcade solution of Eq.~(\ref{helm}) has the following magnetic field components
\begin{eqnarray}\label{bxp} B_x(x,z)&=&B_0\frac{l}{k}\cos{k x}\, e^{-l\, z},\\
B_y(x,z)&=&B_0\frac{\alpha}{k}\cos{k x}\, e^{-l\, z},\\
B_z(x,z)&=&-B_0\sin{k x}\, e^{-l\, z}.\label{bzp} \end{eqnarray}

\noindent where $B_0$ is a reference constant. The parameter $k$ is related to
the lateral extension of the arcade while $l$ is a measure of the vertical
magnetic scale height. The parameter $\alpha$ is given by the following
expression 
\begin{eqnarray}\label{alpha} 
\alpha=\left(k^2-l^2\right)^{1/2},
\end{eqnarray}
and it is related to the amount of shear in the arcade. For $l=k$, $\alpha=0$, the
magnetic field is purely potential and the $B_y$ component is zero.

Magnetic field lines in the configuration given by Eqs.~(\ref{bxp})-(\ref{bzp})
do not show any dips because the configuration is bipolar. A simple way to
obtain a configuration with magnetic dips  it to chose a particular superposition
of two magnetic arcades representing a quadrupolar configuration 
\begin{eqnarray}\label{bx} B_x(x,z)&=&B_1\frac{l_1}{k_1}\cos{k_1 x}\, e^{-l_1\,
z}-B_2\frac{l_2}{k_2}\cos{k_2 x}\, e^{-l_2\, z},\\
B_y(x,z)&=&B_1\frac{\alpha}{k_1}\cos{k_1 x}\, e^{-l_1\,
z}-B_2\frac{\alpha}{k_2}\cos{k_2 x}\, e^{-l_2\, z},\\
B_z(x,z)&=&-B_1\sin{k_1 x}\, e^{-l_1\, z}+B_2\sin{k_2 x}\, e^{-l_2\, z}.\label{bz} \end{eqnarray}
The individual arcades, quoted with the sub-indexes 1 and 2, must have the same
$\alpha$ to satisfy the Helmholtz equation given by Eq.~(\ref{helm}). Thus, we have according to Eq.~(\ref{alpha})
 the constraint
\begin{eqnarray}\label{alphac} 
k_1^2-l_1^2=k_2^2-l_2^2.
\end{eqnarray}
The width of the full structure is $L_a$, and we select the following wavenumbers
$k_1=\pi/2L_a$ and $k_2=3\pi/2L_a$ since they generate a magnetic configuration
with dips at $x=0$. The parameter $l_1/k_1$, hereafter referred as $l/k$, is
related to the amount of shear in the structure. Several examples of different sheared
arcades are found in Fig.~\ref{sheardep} by imposing the previous wavenumbers
and that $B_2=B_1$. Note the location of dips and
that the magnetic configuration is invariant
in the $y-$direction.

The initial prominence mass and size are prescribed according to typical values reported
from observations \citep[see for example][]{labrosseetal10}. Here we assume that the
typical size of the density enhancement has a width of $5\times 10^3\,\rm km$, a length of
$4\times 10^4\,\rm km$, and a height of $10^4\,\rm km$. The prominence is initially
suspended above the photosphere at a height of  $1.75\times 10^4\,\rm km$. With a typical
prominence density of $5.2 \times 10^{-2}\,\rm kg\, km^{-3}$ and taking into account the
geometry of the prescribed prominence (see Fig.~\ref{promva20side_1} top panel), the total
mass is $1.3 \times 10^{11}\,\rm kg$. For this case the density contrast between the
corona and the core of the prominence is around 120. In some simulations we have changed
the total mass of the prominence by reducing the density contrast. Between the core of the
prominence and the corona we have included in the initial density profile a
prominence-corona transition region (PCTR) with a typical width of $30-15\%$ of the
characteristic length in each direction. The mass deposition is produced just at
$t=0$ and contrary to the situation in \citet{terradasetal2013} the deposition is
instantaneous.

Note that there is no initial velocity perturbation introduced in the system.
However, because of the mass deposition at $t=0$ the system immediately reacts to the
presence of the enhanced mass which is pushed down by the gravity force, and therefore
velocity perturbations, specially in the vertical component, will be generated.

\section{Numerical tools}\label{numtools}

The code used to solve the ideal MHD equations is an evolution of the code MoLMHD
already described in \citet{terradasetal2013} \citep[see also][]{bonaetal09}. The main
novelty in the present version is the implementation of a WENO scheme \citep[see for
example the review of ][]{shu2009} to solve some of the MHD equations
\citep[see][]{jiangwu1999,balsaraetal2009}. In particular we have used a fifth order
accurate method in space to solve the equations of continuity, momentum and energy. This
last equation has been written in terms of gas pressure instead of total energy. The
WENO scheme is very robust with respect to the presence of stiff gradients in the
variables and thus suitable to handle strong shocks. The density contrast between the
prominence and corona can be higher than 100, and therefore difficult to treat without
high-resolution numerical methods. Nevertheless, for the induction equation we have used
a standard fourth order central discretization with an artificial dissipation
coefficient since the application of the WENO scheme on the magnetic field has been
found to be too diffusive. The excess of diffusion produces plasma motions across  the
magnetic field changing the dynamics with respect to the idealized case where plasma is
frozen to the magnetic field. The inevitable numerical diffusion associated to the
scheme has been significantly reduced with the method used in this paper.  

 The MHD equations solved in this study are
\begin{eqnarray*}
\frac{\partial{\rho}}{\partial t}+\nabla\cdot \left({\rho \bf v}\right) =0,
\end{eqnarray*}

\begin{eqnarray*}
\frac{\partial{\rho \bf v}}{\partial t}+\nabla\cdot \left({\rho \bf v \bf v}
+p {\bf I}-\frac{\bf B \bf B}{\mu}+\frac{{\bf B}^2}{2 \mu}
\right) =\bf \rho g,
\end{eqnarray*}

\begin{eqnarray*} \frac{\partial{\bf B}}{\partial t}=\nabla \times \left(\bf v
\times \bf B \right),   \end{eqnarray*}

\begin{eqnarray*}
\frac{\partial{p}}{\partial t}+\nabla\cdot \left({\gamma p \bf v}\right) =
\left(\gamma-1\right) {\bf v}\cdot \nabla p,
\end{eqnarray*}

\noindent where $\bf I$ is the unit tensor, $\bf g$ is the gravitational acceleration, and
the rest of the symbols have their usual meaning.

As in \citet{terradasetal2013} the technique of background splitting for the magnetic
field  \citep[see][]{powelletal99} has been used and this has allowed us to gain numerical
accuracy. The splitting introduces source terms in the equations that have been treated
using the fourth order central scheme. Although in  \citet{powelletal99} it is assumed that
the initial magnetic field is potential, i.e., current free, it turns out that for a
magnetic field that is not potential the corresponding source terms are exactly the same.

Regarding boundary conditions we have applied line-tying  at the bottom plane,
representing the photosphere. This condition means that the three components of the
velocities are set to zero while the normal component of the magnetic field,  $B_z$ in our
Cartesian system, is fixed. For the rest of the variables at this boundary we have used
simple extrapolation, i.e., that their spatial derivatives are zero. For top and
lateral boundaries we expect the energy to escape the system. We have tested different
conditions. Using decomposition in characteristics fields  at the boundary
\citep[see][]{terradasetal2013} works fine but it is computationally expensive. We have
found that using fixed values at the boundary essentially produce the same results as the
decomposition in characteristics, and it is much more simple to implement. This last
approach has already been used in \citet{torokkliem2003,toroketal2004} in the context of ideal
kink instabilities in magnetic loops.

The reference length in the simulations is $L=10^4\,\rm km$ and the width of the
arcade is $L_a=6\,L$. Velocities have been normalized to the coronal sound speed
at a temperature of $1\,\rm MK$, $c_{s0}=166\,\rm km\,s^{-1}$. The reference
time is $\tau=L/c_{s0}=1\,\rm min$. The total time of evolution is $100\,\rm
min$. The typical resolution that we have used in the simulations is
$150\times150\times100$ points, with $-6<x/L<6$, $-6<y/L<6$, and  $0<z/L<8$,
being $L=10^4\,\rm km$ the reference length. Higher resolutions have been also
considered to see the effect on the results. Since we consider a global
prominence model we need, from the numerical point of view, to have enough grid
points in both the coronal part and the prominence body. The typical distance
between grid points is $800\,\rm km$ but in some cases it has been reduced to
$550\,\rm km$. Structures of the size of individual threads are not resolved in
this work.

\section{Results}

\subsection{A typical run}\label{0run}

\begin{figure}[!hh] \center{\includegraphics[width=6.5cm]{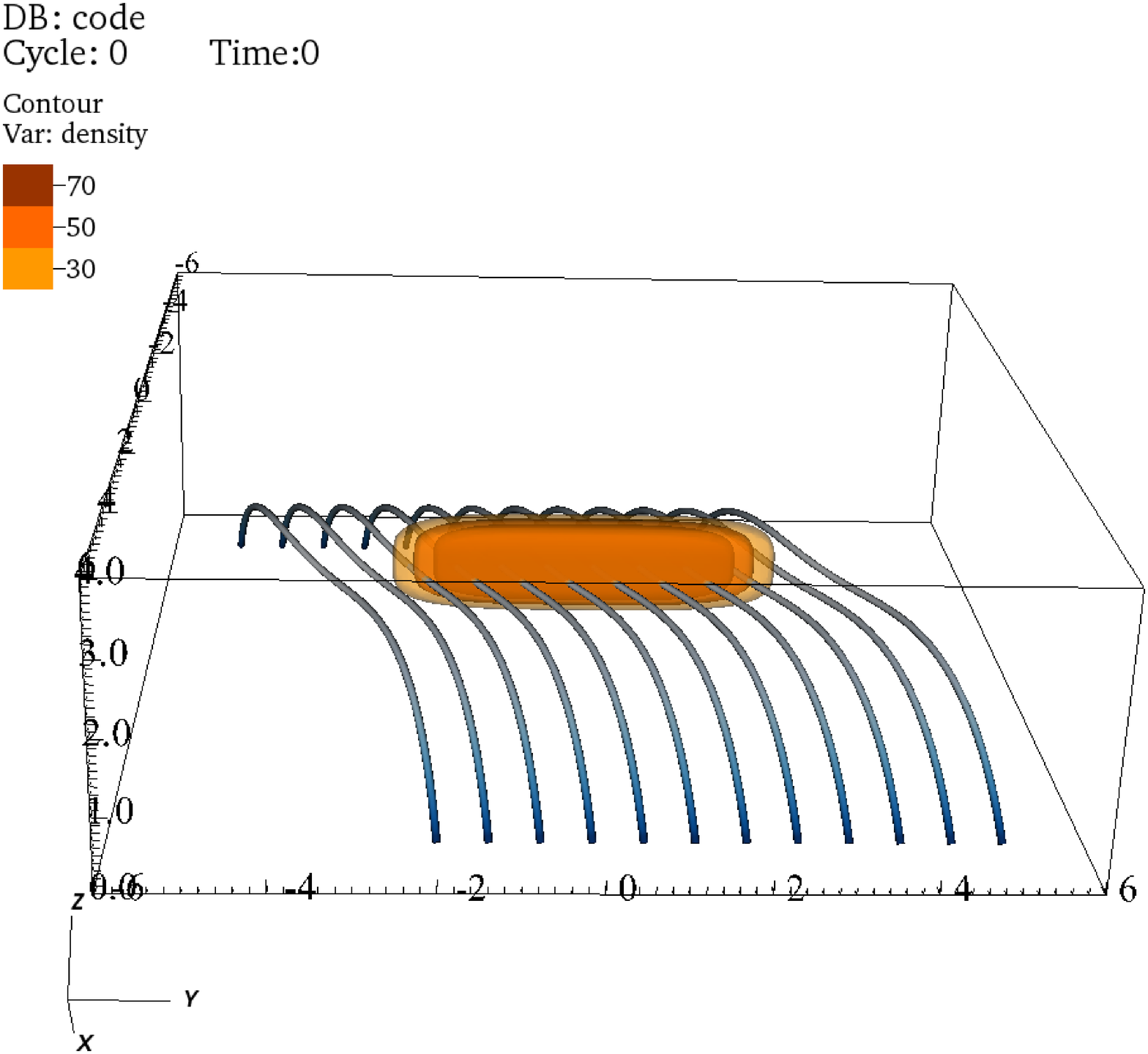}
\includegraphics[width=6.5cm]{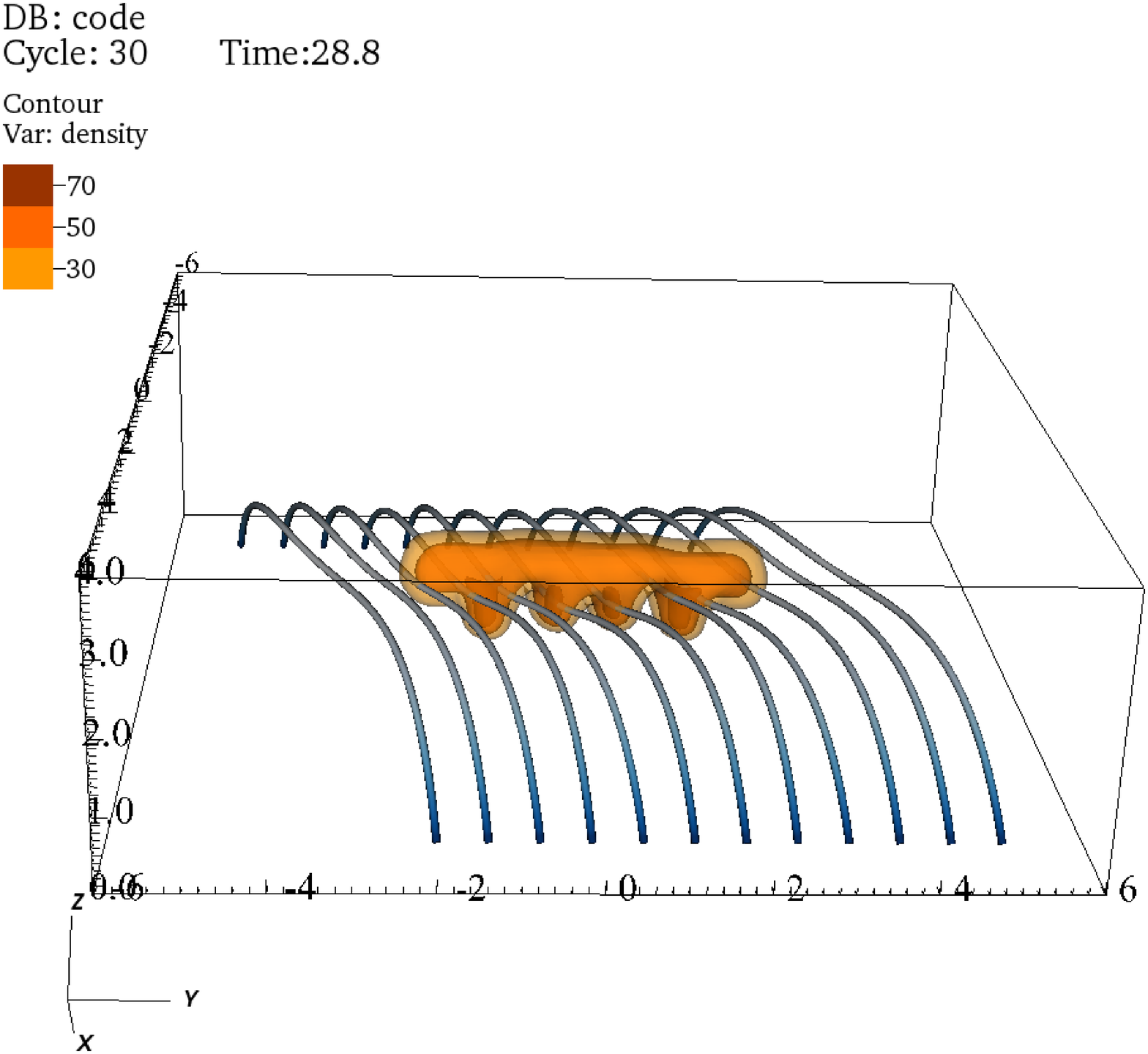}
\includegraphics[width=6.5cm]{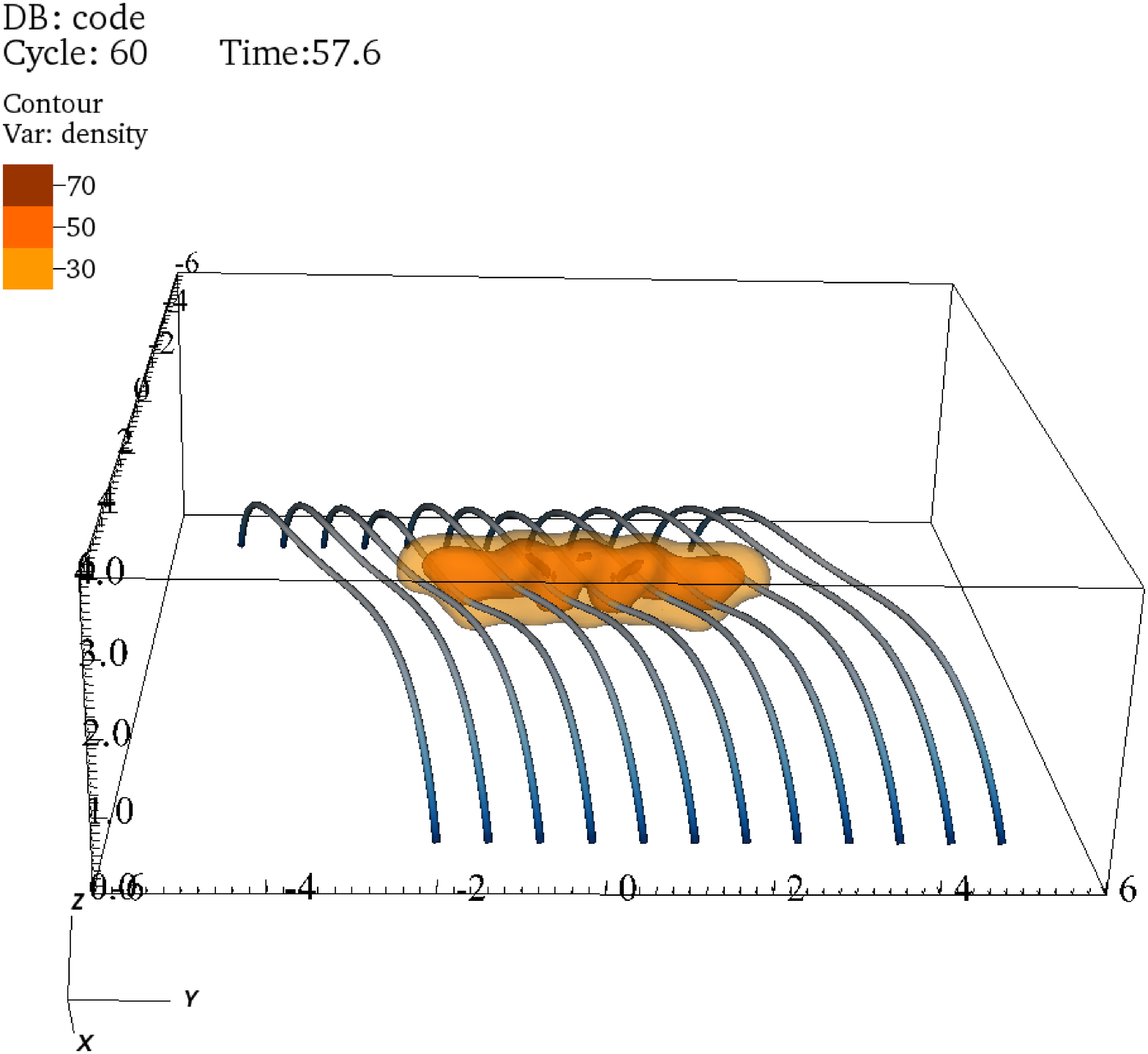} } \caption{\small Time
evolution of density and  magnetic field lines for a typical case. In this
simulation $l/k=0.95$ (weak shear), and $v_{A0}=20\, c_{s0}$, being
$v_{A0}=B/\sqrt{\mu \rho_0}$ the maximum Alfv\'en speed in the system. The
maximum magnetic field ($B$) corresponds to the points $x=\pm L_a$ and $z=0$. Density is
normalized to the coronal density ($\rho_0$). Time in minutes is shown at the
top of each panel. See also Movie1 in the online
material.}\label{promva20side_1} \end{figure}

The results of a typical simulation are displayed in Fig.~\ref{promva20side_1}. In this
figure the density distribution and some specific magnetic field lines are plotted as a
function of time (see also Movie1 in the electronic version of the paper). During the
first hour of evolution the prominence body is able to be suspended above the
photosphere due to  the force provided by the magnetic field. The structure
resembles a detached prominence, referred in the past as a suspended cloud. Although
there is global support, the shape of the prominence changes with time. We can see that
for example after $30\,\rm min$ of evolution, small scales appear in the density
distribution which was initially quite homogeneous. This is linked to the excitation of
MRT instabilities.  Changes in density involve variations of the magnetic field since
the plasma is frozen to the magnetic field in ideal MHD. Therefore the motion of
magnetic field lines has to be consistent with plasma motions (see Movie2).

\begin{figure}[!hh] \center{\includegraphics[width=6.5cm]{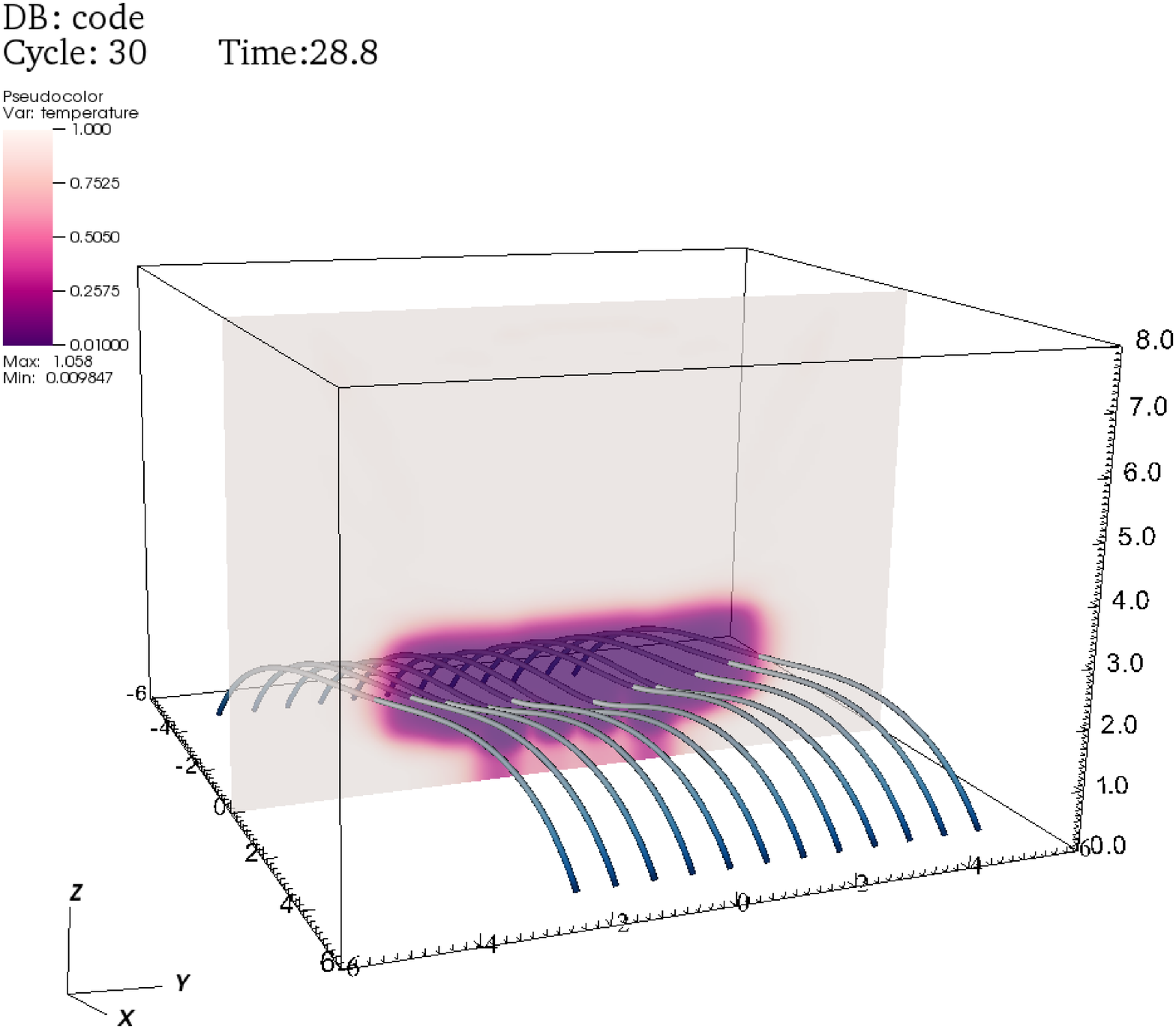}
\includegraphics[width=6.5cm]{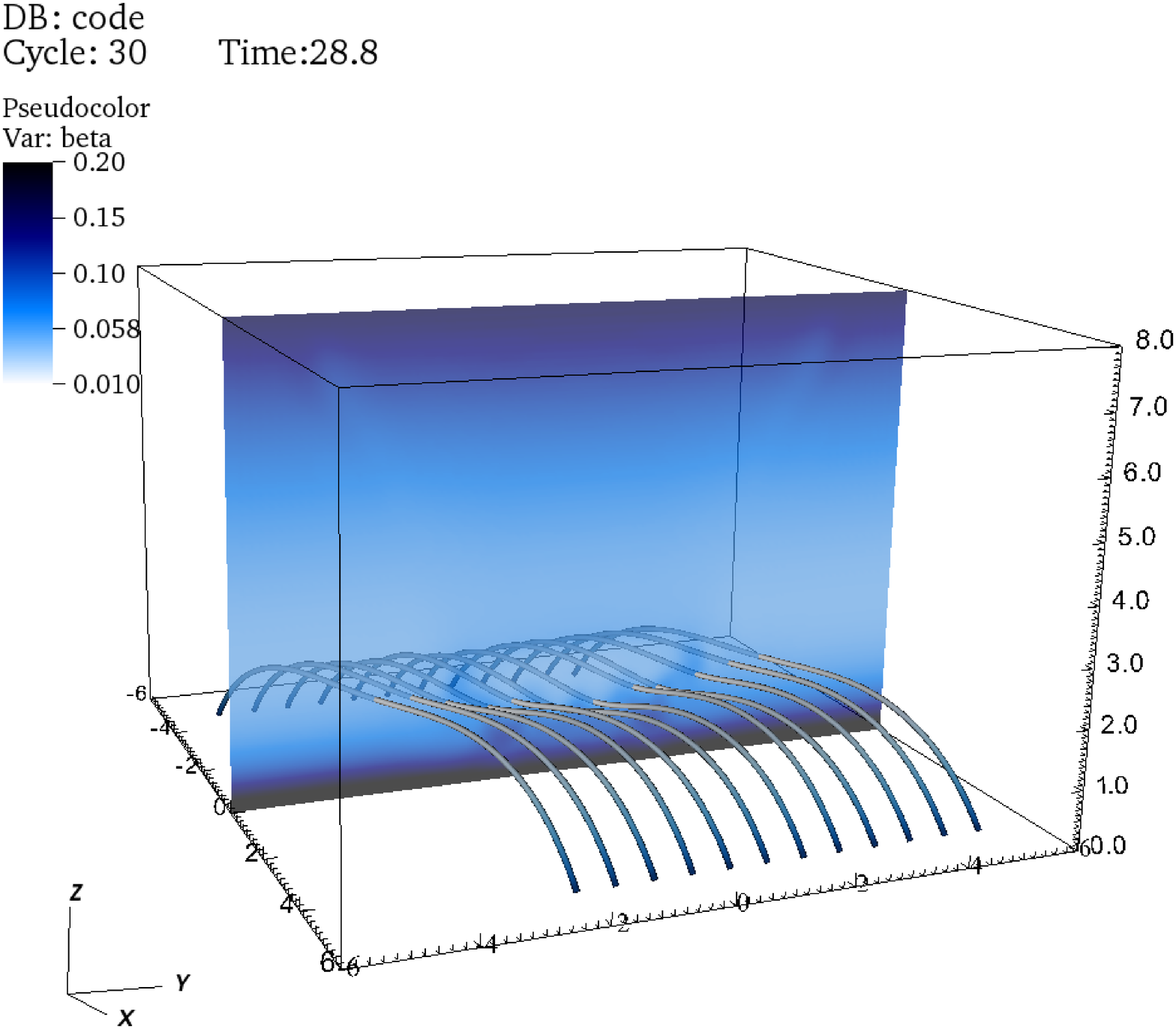}} \caption{\small Temperature and plasma-$\beta$ at
the plane passing through $x=0$ for the simulation shown in Fig.~\ref{promva20side_1} at
$t=28.8\,\rm min$. Temperature is normalized to the coronal value, $1\,\rm
MK$.}\label{promva20betemp} \end{figure}

At a given instant, temperature and plasma$-\beta$ distributions ($\beta$ is the
coefficient between gas and magnetic pressure) are quite irregular due to the presence
of the cool and dense material representing the prominence. The distribution of these
magnitudes at the central plane is plotted in Fig.~\ref{promva20betemp}. We find that
density variations produced during the evolution also involve temperature changes. 
Temperature is typically of the order of $10\,000\,\rm K$ inside the prominence while in
the corona is around  $1\,\rm MK$. Between these two regions we find the PCTR, clearly
visible in temperature. In this configuration  the plasma$-\beta$ changes mostly with
height, it has a minimum around the prominence location and increases toward the
photosphere and upwards with height. At the the prominence core it is below 0.02.

\begin{figure}[[!hh] \center{\includegraphics[width=6.5cm]{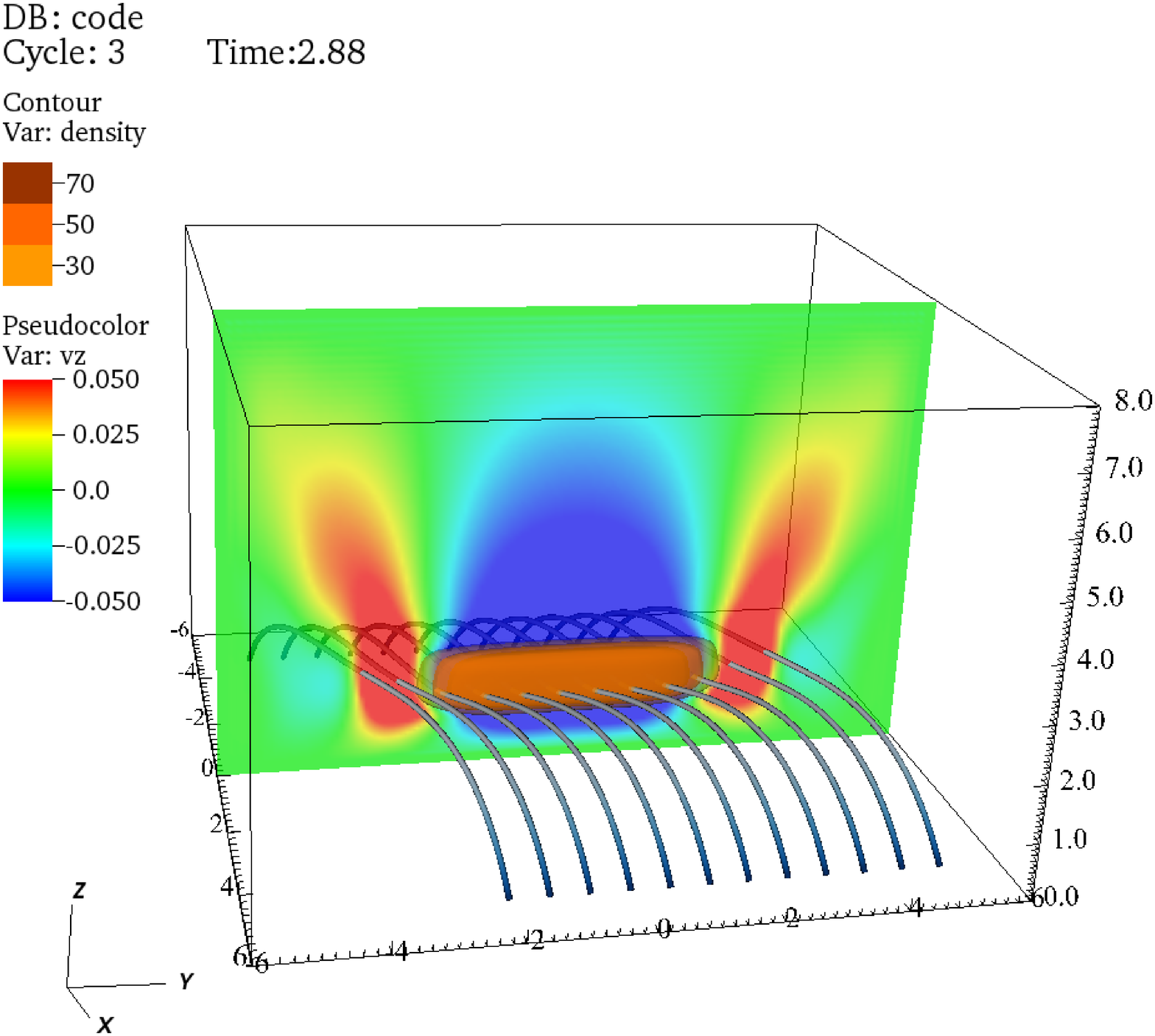}
\includegraphics[width=6.5cm]{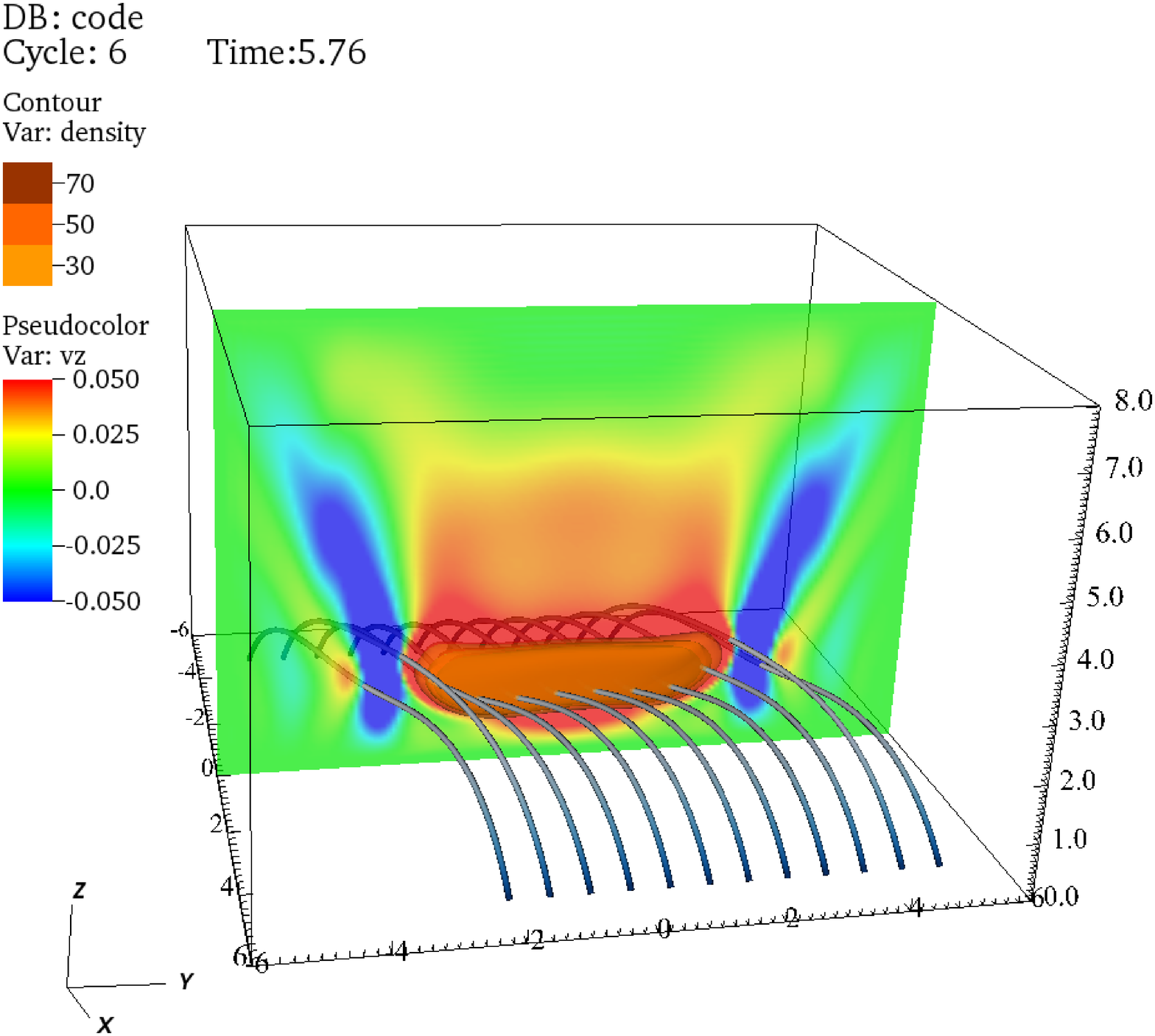}
\includegraphics[width=6.5cm]{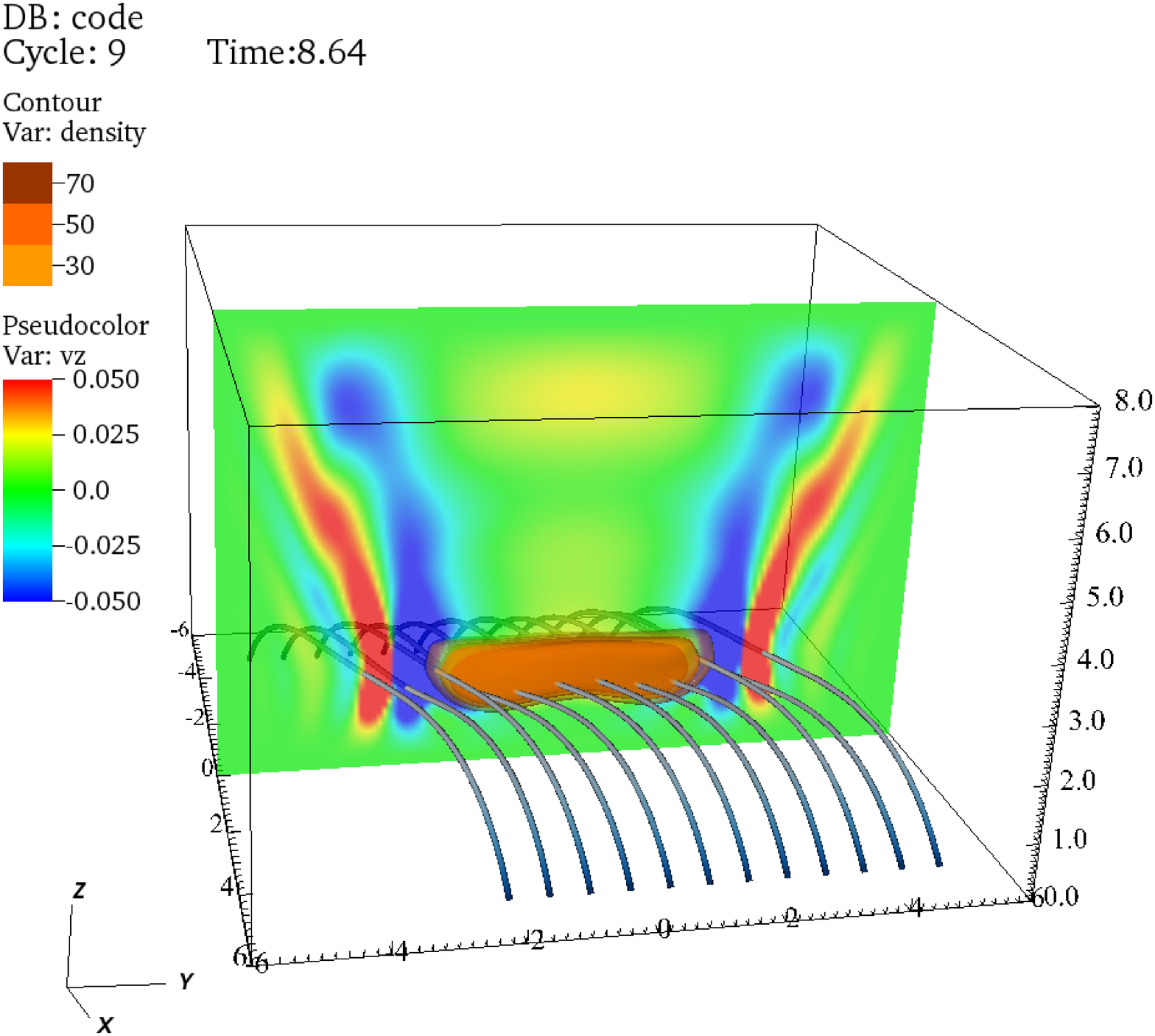}} \caption{\small Time evolution
of  density and vertical velocity at the $yz-$plane at $x=0$ during the first
$12 \,\min$ of evolution before the onset of the instability. Shear vertical
motions are produced at the boundaries of the prominence during the
oscillations. Velocity amplitudes are normalized to the coronal sound speed.
}\label{promva20vel} \end{figure}

The dynamics of the system is quite rich and several features need to be described in
detail. For example, the prominence shows oscillations before the clear onset of the MRT
instability. In  Fig.~\ref{promva20vel} the vertical component of the velocity at early
stages of the simulation is represented at the plane passing through the prominence
core. At the center of the prominence the velocity at  $t=2.88\min $ is negative since
the prominence body is moving downwards. In the following frame, at $t=5.76\min $ the
motion is in the positive vertical direction. This is a consequence of the vertical
oscillatory motion that the whole prominence is doing during the relaxation process.
Note also the deformation of the lower boundary of the prominence displaying a convex
curve for $t=8.64\min$. At the same time strong shear motions are generated at the
lateral edges of the prominence where we find positive/negative velocity patterns in
$v_z$. The spatial scales of these patterns decrease with time as can be appreciated in
Fig.~\ref{promva20vel}. This has to do with the process of mode conversion and
phase-mixing that takes place at the inhomogeneous layers of the structure, in this case
the prominence PCTR at the sides of the prominence body in the $y$-direction. Part of
the attenuation of the oscillatory vertical motion of the structure is due to this
process, however, wave leakage might play a role as well. Since our aim is to
concentrate on the global evolution of the system we leave the analysis of oscillations
and damping for future studies. However, it is interesting to mention that these shear
motions at the edges of the prominences may lead to the development of another type of
instabilities due to velocity shear, i.e., Kelvin-Helmholtz  instabilities. Indeed, a
closer inspection of the lateral edges of the prominence indicate the presence of very
dynamic motions at the PCTR associated to this kind of instability, but with the
particularity that shear changes periodically with time due to the global oscillation of
the structure. This effect is similar to the deformations produced at the boundaries of
a cylindrical tube when it is oscillating \citep[see][]{terradasetal08}.

\begin{figure}[!hh] \center{\includegraphics[width=6.5cm]{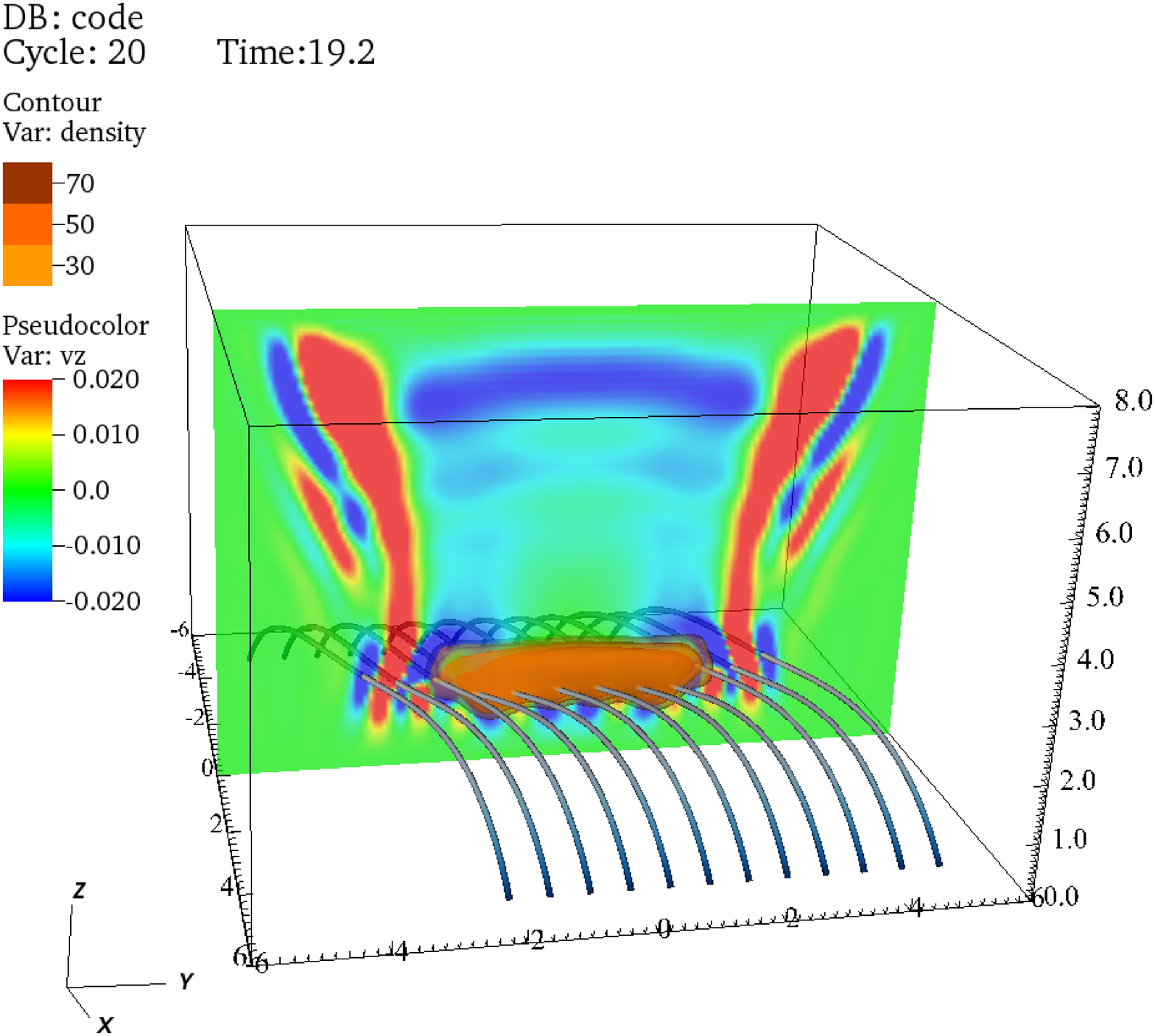}
\includegraphics[width=6.5cm]{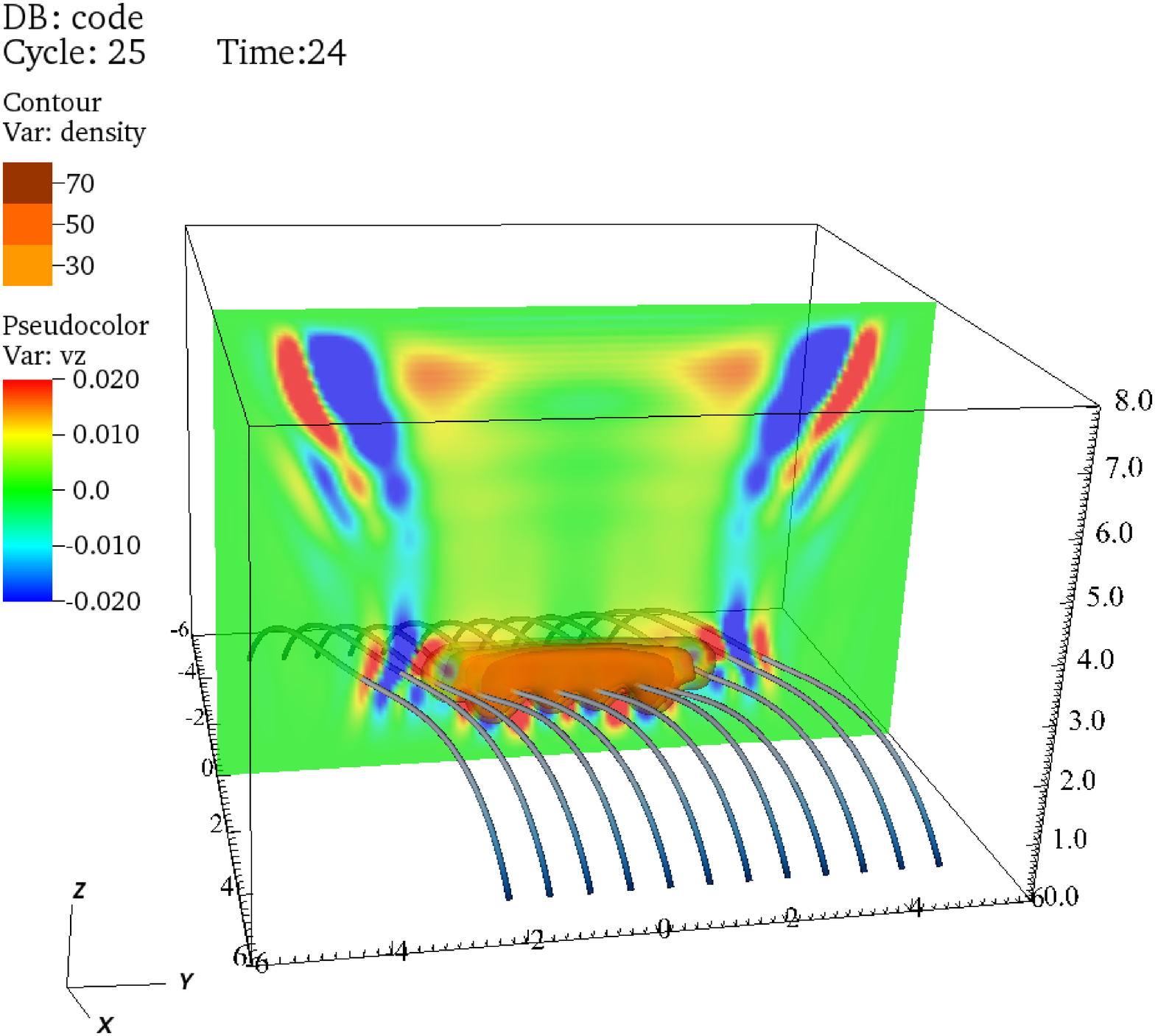}
\includegraphics[width=6.5cm]{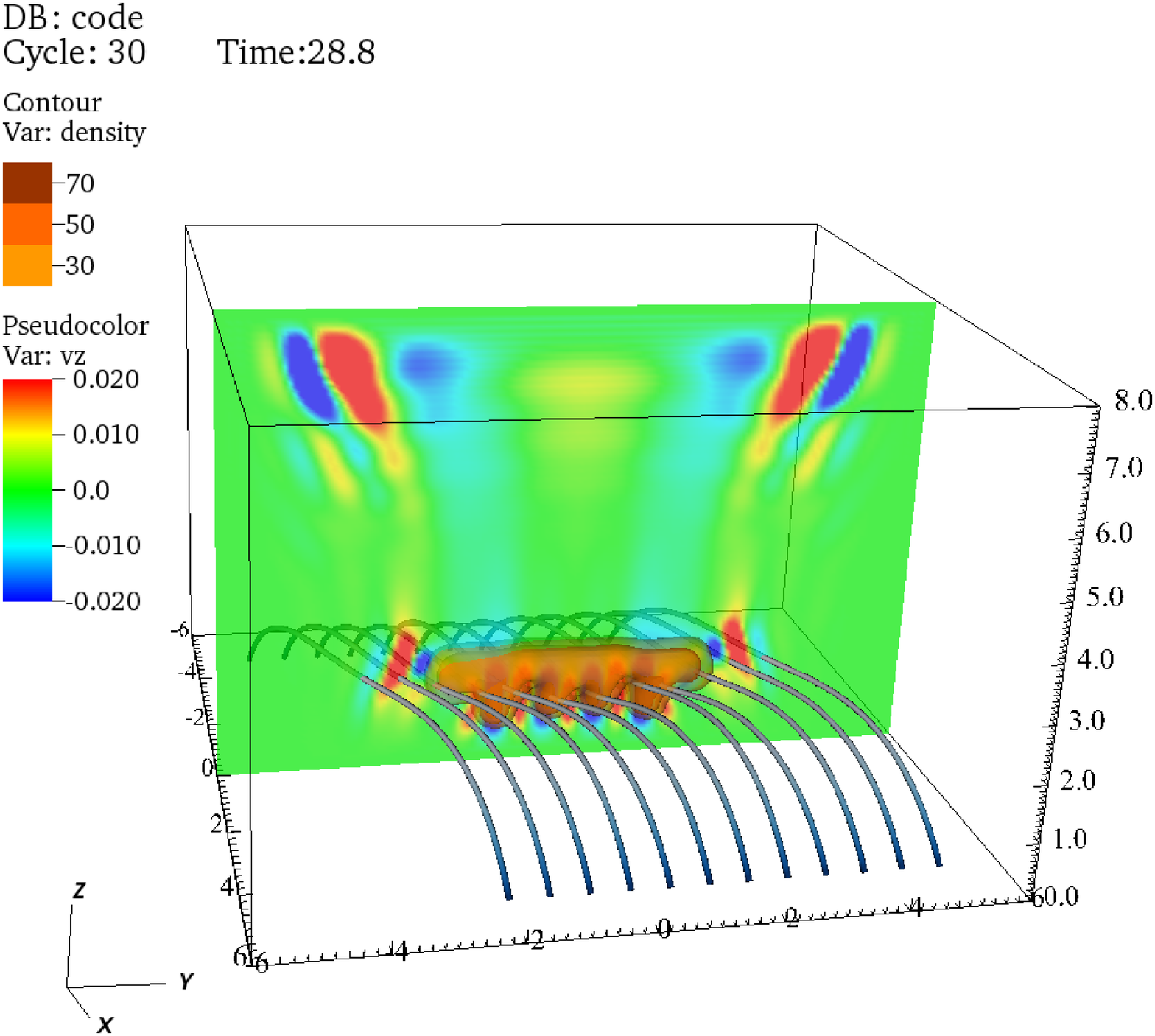}
} \caption{\small Time evolution of 
density and vertical velocity at the $yz-$plane at $x=0$ during the development of the 
MRT instability.}\label{promva20vel1} \end{figure}

\begin{figure}[!hh] \center{\includegraphics[width=6.5cm]{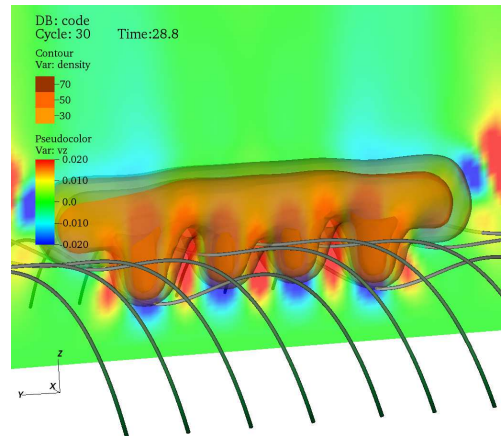}} \caption{\small
Zoom of density and vertical velocity at a given time (see Fig.~\ref{promva20vel1}
bottom panel). Fingers and plumes associated to the MRT instability are present and are
associated to the shear motions in velocity. Some magnetic field
lines are also represented. See Movie2 in the online material.}\label{densbig}
\end{figure}

\begin{figure}[!hh] \center{\includegraphics[width=8cm]{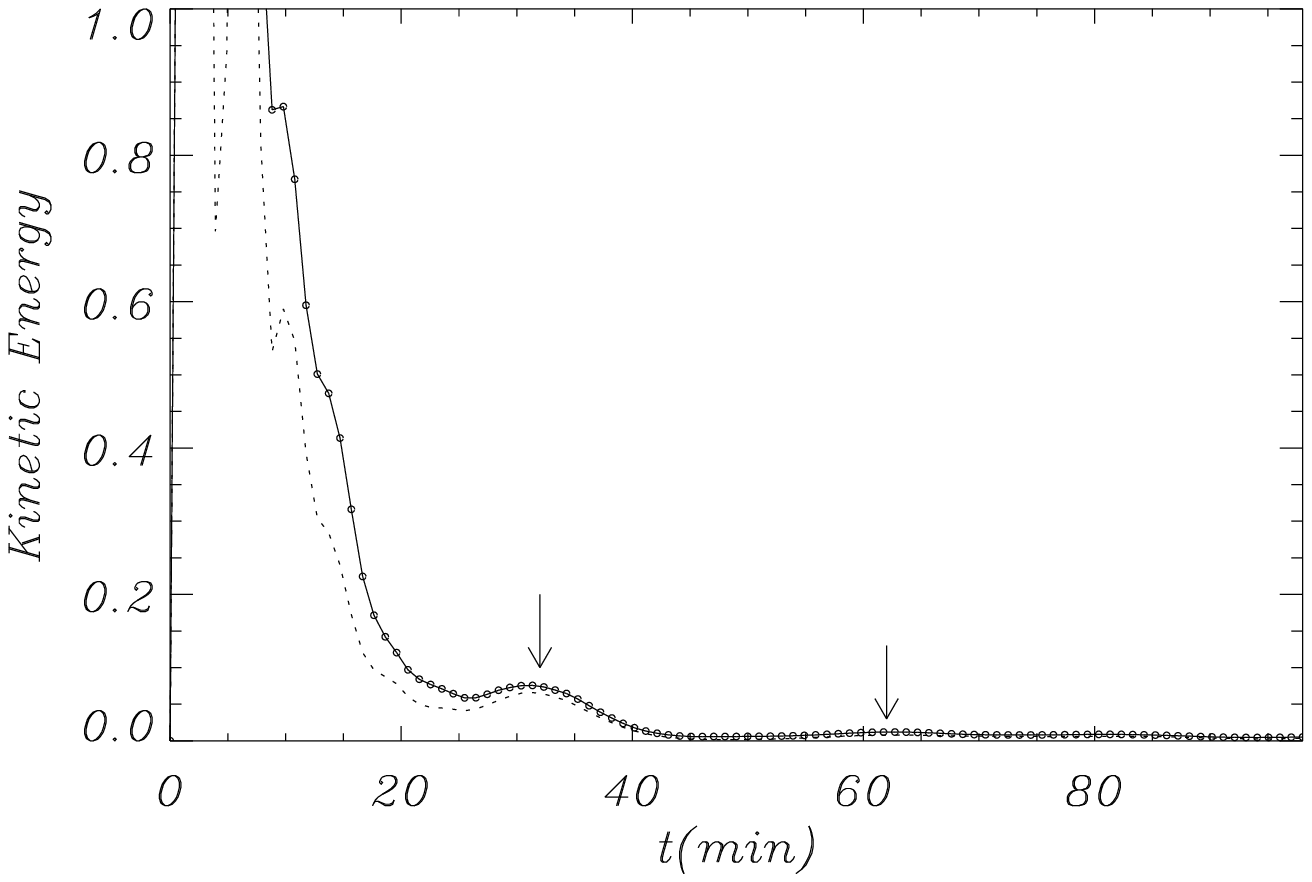}} \caption{\small 
Integrated kinetic energy at the $yz-$plane at $x=0$ as a function of time. The continuous
line represents the total kinetic energy while the dotted line corresponds to the kinetic
energy in the vertical direction only. The arrows show times when the MRT
instability is clearly identified.}\label{kinetics}
\end{figure}

At the beginning of the simulation there are no clear signatures of MRT unstable
modes. Instabilities start to develop around $t=20\,\rm min$, as can be
appreciated in Fig.~\ref{promva20vel1}. Apart from the shear motions associated
to the lateral edges of the prominence and explained before, we also find shear
motions in the vertical component of the velocity that start to form at the
bottom part of the prominence (see positive/negative velocity patterns). These
vertically shear motions now produce strong deformations of the bottom of the
prominence, clearly visible in density in Fig.~\ref{promva20vel1}. Descending
fingers of cool and dense plasma and ascending plumes of hot and less dense
plasma have been developed in the structure and modify significantly the density
distribution of the prominence. For the particular simulation studied here the
prominence bottom shows several arches that spread all over the prominence body.
In particular we can identify up to four clear fingers. The density and $v_z$
distribution is further visualized in Fig.~\ref{densbig} (see also Movie2). The
excitation of MRT unstable modes affects the prominence on a global scale and
the instability is modified by the effect of line-tying conditions. For the
simulation shown in Figs.~\ref{promva20vel1} and ~\ref{densbig} the growth time
of the instability is of the order of $20\,\rm min$. This estimation is based on
the change in position of the finger-like structures in the prominence but it is
important to remark that during the evolution of the MRT instability the
prominence is still oscillating (see Movie2) and this complicates the estimation
of growth-rates. Thus, this growth time does not correspond to the growth time
that one would calculate in the linear stage of development of the instability.
Nevertheless, it is a useful parameter since it gives a typical time scale
associated to the physical process. It is worth to mention that in this
simulation the prominence is not destroyed by the instability at least during
the first $100\,\rm min$ of evolution. The eventual destruction of the
prominence in a short period of time would mean that indeed a continuous supply
of material is required as some observations suggest.

Further information about the dynamics of the system is derived from the behavior of
the energy. We have calculated from the simulations the integrated kinetic energy at the
plane $yz-$plane ($x=0$). This magnitude is useful since it shows clear signatures of the
development of instabilities associated to MRT unstable modes. In  Fig.~\ref{kinetics} the
total kinetic energy (continuous line) and the energy associated to the vertical velocity
component (dotted line) are plotted as a function of time. According to the plot, most of
the kinetic energy is due to vertical motions which are initially induced by the gravity
force. The energy decreases with time due to the effect of leakage while the prominence is
oscillating vertically (see the fluctuations in the dotted line for $t<20\,\rm min$).
Later, around $t=25\,\rm min$ the kinetic energy starts growing, this is the indication of
the development of the instability and coincides with the appearance of the fingers and
plumes in density (see for example, Fig.~\ref{promva20side_1} middle panel, or Movie1). After peaking around $t=32\,\rm min$ the energy starts decreasing again, meaning that
fingers do not continue moving toward the photosphere at the same velocity, in fact they
are decelerating. In the energy profile (and this can be also identified in Movie1) there
is a faint indication about a new development of the instability around $t=60\,\rm min$.
Therefore in this particular example the instability is unable to destroy the prominence
although it produces dynamic motions at the prominence body.

The MRT unstable interface between the prominence and corona has smooth density gradients
due to the presence of the PCTR. It is well known that density gradients have the effect
of reducing the growth rate of RT instabilities, especially at short wavelengths
\citep[see for example][]{mikaelian1986}. We have performed a run with half the size
of the PCTR in the standard simulation but the differences in the spatial scales of the
instability and the growth rates are not too significant. Although we start with a given
width of the PCTR, after the time-evolution the eventual ``quasi-equilibrium" that is
obtained has a thicker transition between the core of the prominence and corona since
radiation and conduction are neglected in the model. The
magnetic field also has gradients and  \citet{yangetal2011} have shown that they affect
the linear growth rates of the MRT instability in a certain range that depends on the
magnetic field strength. Thus, the excitation of the most unstable mode with a particular
spatial scale, in our case we find structures with typically three cavities,
will be conditioned by all these effects and it is very difficult to anticipate. 
However, we think that the most important factors that affect the appearance of the
unstable modes and the spatial distribution of the cavities in the horizontal direction
are the strength of the magnetic field at the core of the prominence and the variation of
the shear angle with height. This will be studied in Section~\ref{sheardepsec}. It is
worth to mention that due to the symmetry of the system there is a maximum in velocity at
the center of the prominence (in the $xz-$plane at $y=0$). This means that there is a
natural excitation of symmetric modes in velocity respect to the center of the prominence.
For this reason it is logical to find structures with an odd number of cavities along the
$y-$direction (three in the case of Fig.~\ref{promva20side_1}).

The spatial patterns related to the MRT instability found in the simulations may have a
close link with the observed arches and cavities reported in many prominences, see for
example \citet{bergeretal10}. It is worth to mention that in \citet{bergeretal10} the
cavities beneath the prominence are found forming single structures instead of multiple
cavities like in the present simulations. However, the vertical structuring observed in
many polar crown prominences \citep[see for example][]{dudiketal2012} resembles the
spatial structures found in the simulations. \citet{ryuetal10} associated the appearance
of bubbles/cavities to the development of screw-pinch instabilities assuming helical
prominence models. However, if the structure does not have a clear flux rope shape and it
is better represented by a sheared arcade, then the MRT instabilities found in our
simulations can provide an explanation for the formation of cavities. In this regard, MRT
instabilities have been already investigated using 3D simulations by
\citet{hillieretal11,hillieretal12,hillieretal12a}. Since these authors have focused on
internal motions the connection of the field lines with the photosphere was neglected in
their models, and as we have already anticipated line-tying conditions change considerably
the dynamics of the MRT instability. The MRT instability in our configuration affects the
whole prominence body, instead of small internal parts. However, it must be pointed out
that due to the limitation in the numerical resolution, we are unable to resolve features
with the spatial scales studied by \citet{hillieretal11}. 

\begin{figure}[!hh] \center{\includegraphics[width=6.5cm]{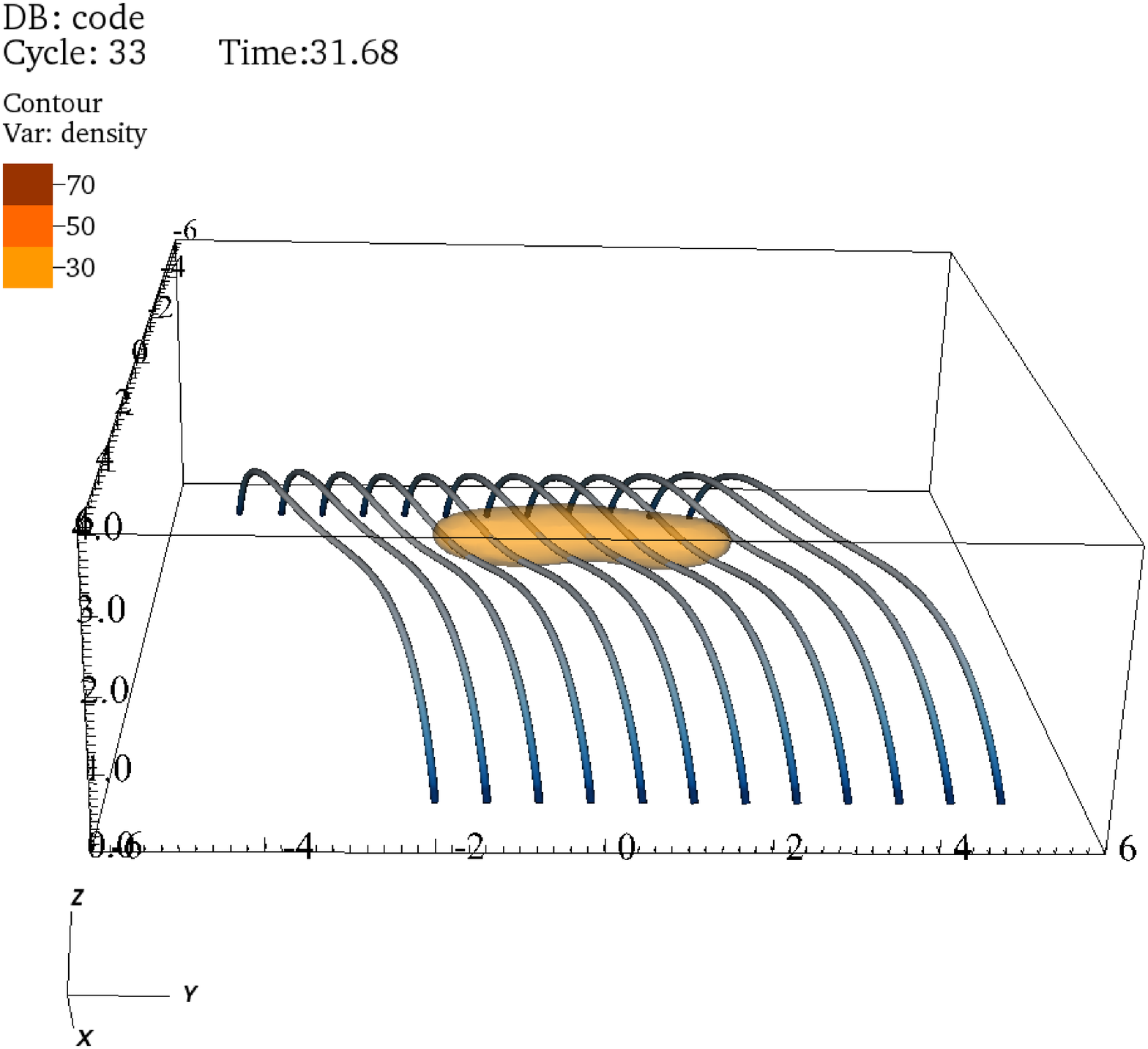}\\
\includegraphics[width=6.5cm]{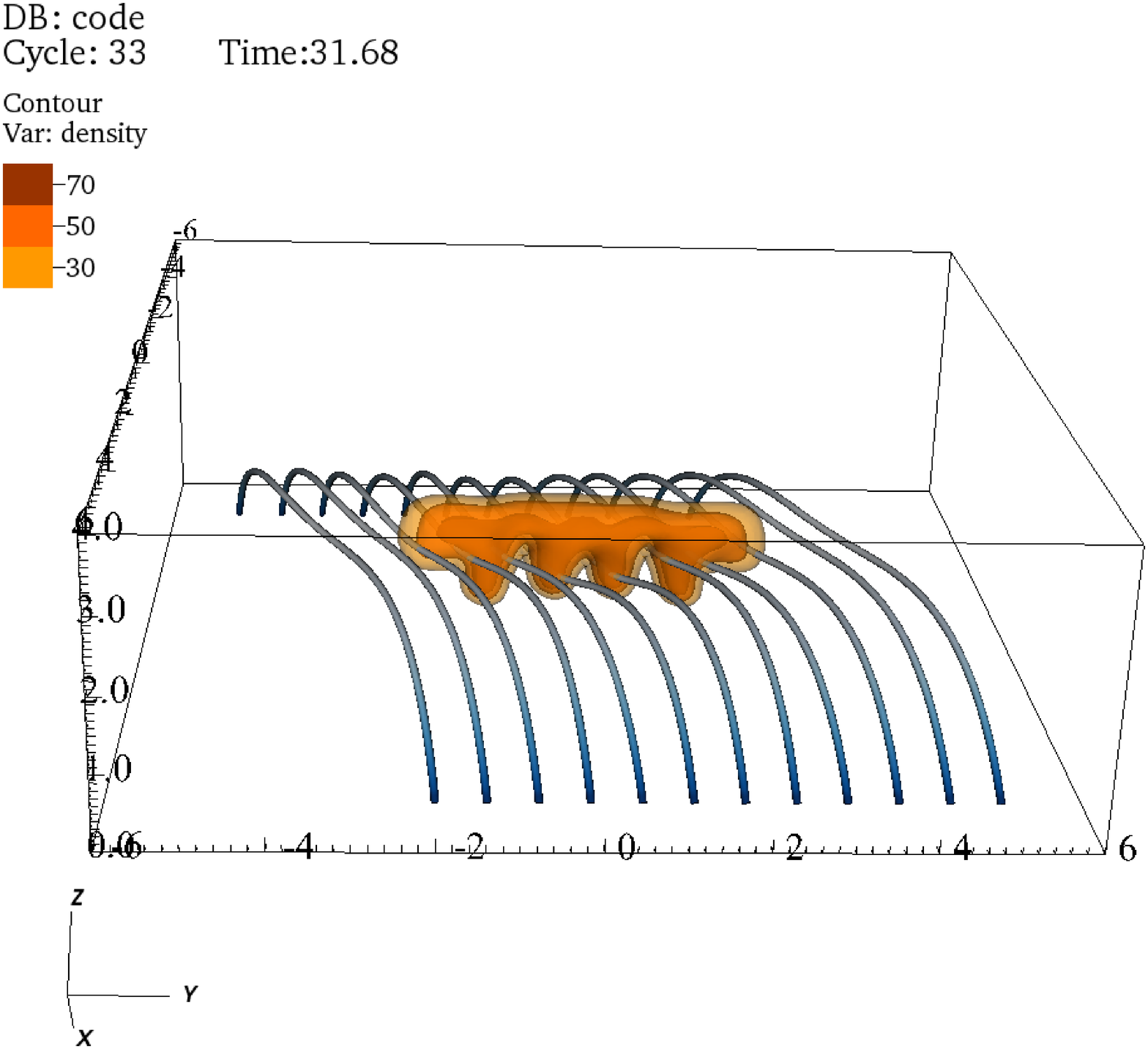}\\ \includegraphics[width=6.5cm]{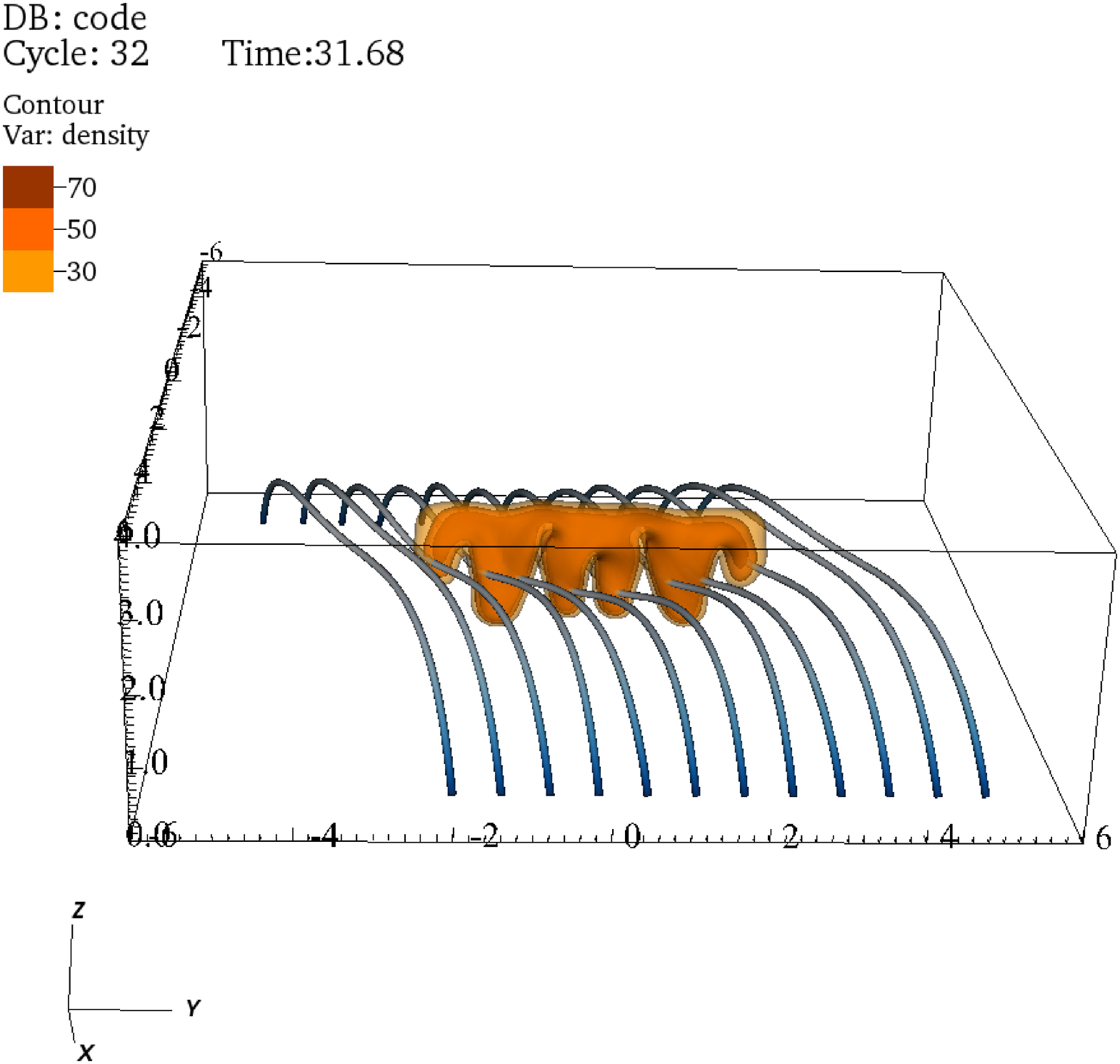} }
\caption{\small Prominence density and magnetic field for three different values of the
numerical resolution, $75\times75\times 50$, $150\times 150\times 100$, and
$225\times225\times 150$. Time is essentially the same for the three simulations. All
the other parameters are the same as in Fig.~\ref{promva20side_1}.}\label{resdep}
\end{figure}

\subsubsection{Effect of boundary conditions}

For the simulations presented here line-tying conditions have been imposed at the
photosphere. Without this condition at $z=0$, the core of the prominence would simply fall
due to the effect of gravity, and the associated changes of the local magnetic field would
be unable to provide magnetic support. We have performed several simulations to check this
effect by removing line-tying photospheric conditions and indeed we have not been able to
find a sustained prominence. Thus, the effect of the communication between the prominence
and the photosphere due to the reflecting conditions cannot be neglected since it is very
relevant regarding the support of the prominence against gravity. This is an important
property of the simulations performed in this work. We plan to study in the future
other boundary conditions such as mirror boundaries or boundaries using potential field
extrapolations. In fact, the photospheric magnetic field may change over timescales of a
few hours due to photospheric convective motions. Nevertheless, we have decided to
concentrate on the most simple boundary conditions that lead to sustained prominences,
i.e., line-tying conditions.

We have also checked the effect of the location of the lateral and upper boundaries. We have
concluded that locating these boundaries at larger distances from the prominence body does
not change significantly the results. For this reason, we have kept the boundaries located
at the distances given in Sect.~\ref{numtools} in order to avoid a significant decrease in
the numerical resolution when a fixed number of grid points is used.

\subsubsection{Effect of numerical resolution}

The complex dynamics of the instability depends on the resolution of the
numerical simulation. To show this effect we have plotted in Fig.~\ref{resdep}
the results at the same instant for three different numbers of grid points. With
a low number of points, although there is still magnetic support, the
instability is simply unresolved and density is too smoothed by the 5th order
WENO scheme (see that high density isocontours are missing). When the resolution
is improved spikes and plumes appear and the instability is resolved. Note also
that with the highest resolution the instability has developed at a faster rate
than at the intermediate resolution. This is due to decrease of the numerical
dissipation when resolution is increased. Along the same lines, it must be
pointed out that the instability will produce a cascade of energy to smaller
spatial scales. If the small scales are not properly resolved this cascade of
energy will not proceed at the appropriate rate and at some point a saturation
will be achieved being the numerical dissipation dominant. This has prevented us
from performing long runs in which the solution is too degraded, and have only
focused on the evolution for about $100\,\rm min$. Future studies should include
nonuniform grids or adaptive mesh refinement (AMR) to resolve spatial scales
smaller than $550\,\rm km$, and to have a better estimation of the growth rates.

\subsection{Dependence on $\beta$}

\begin{figure}[!hh] \center{\includegraphics[width=6.5cm]{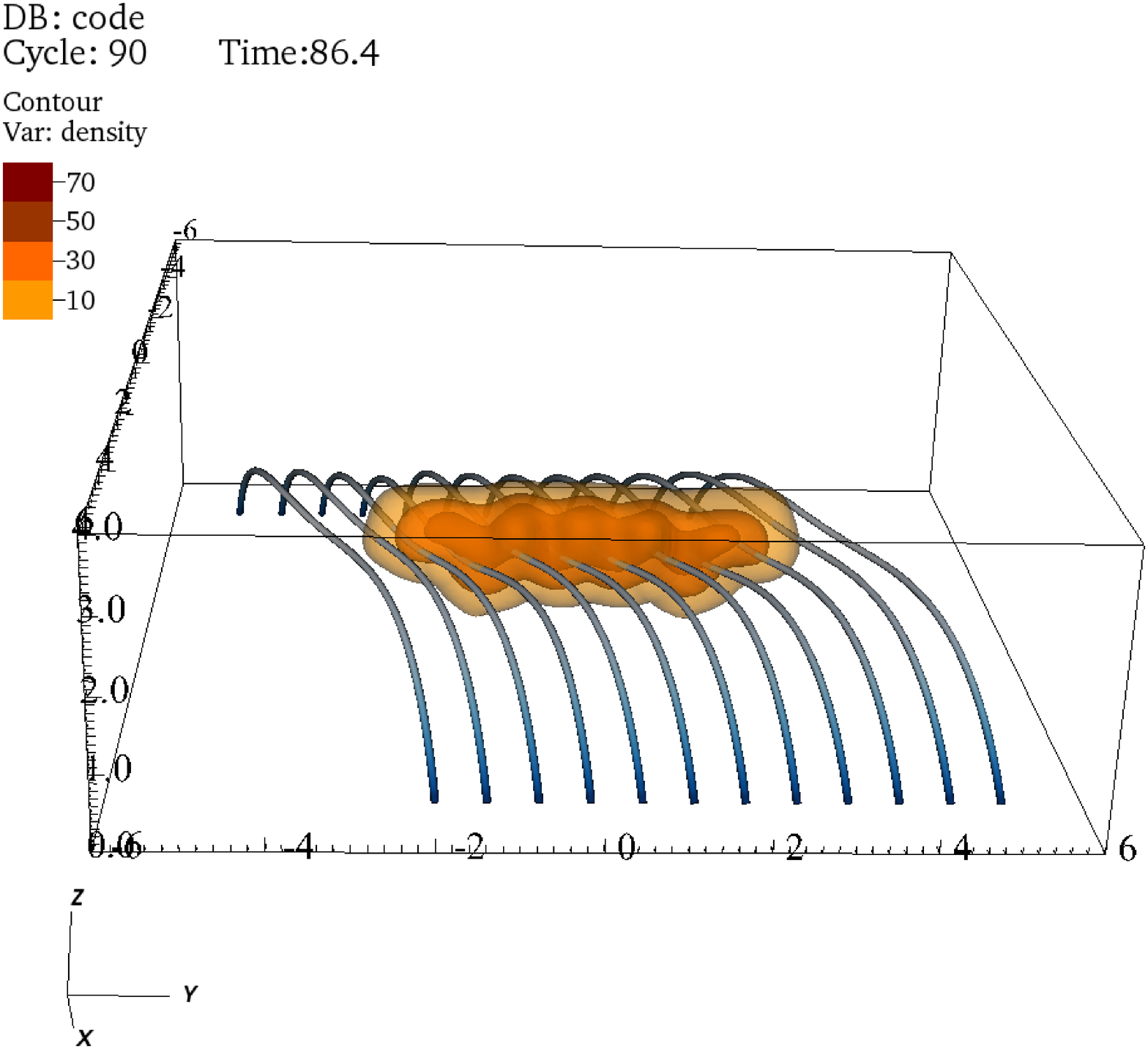}\\
\includegraphics[width=6.5cm]{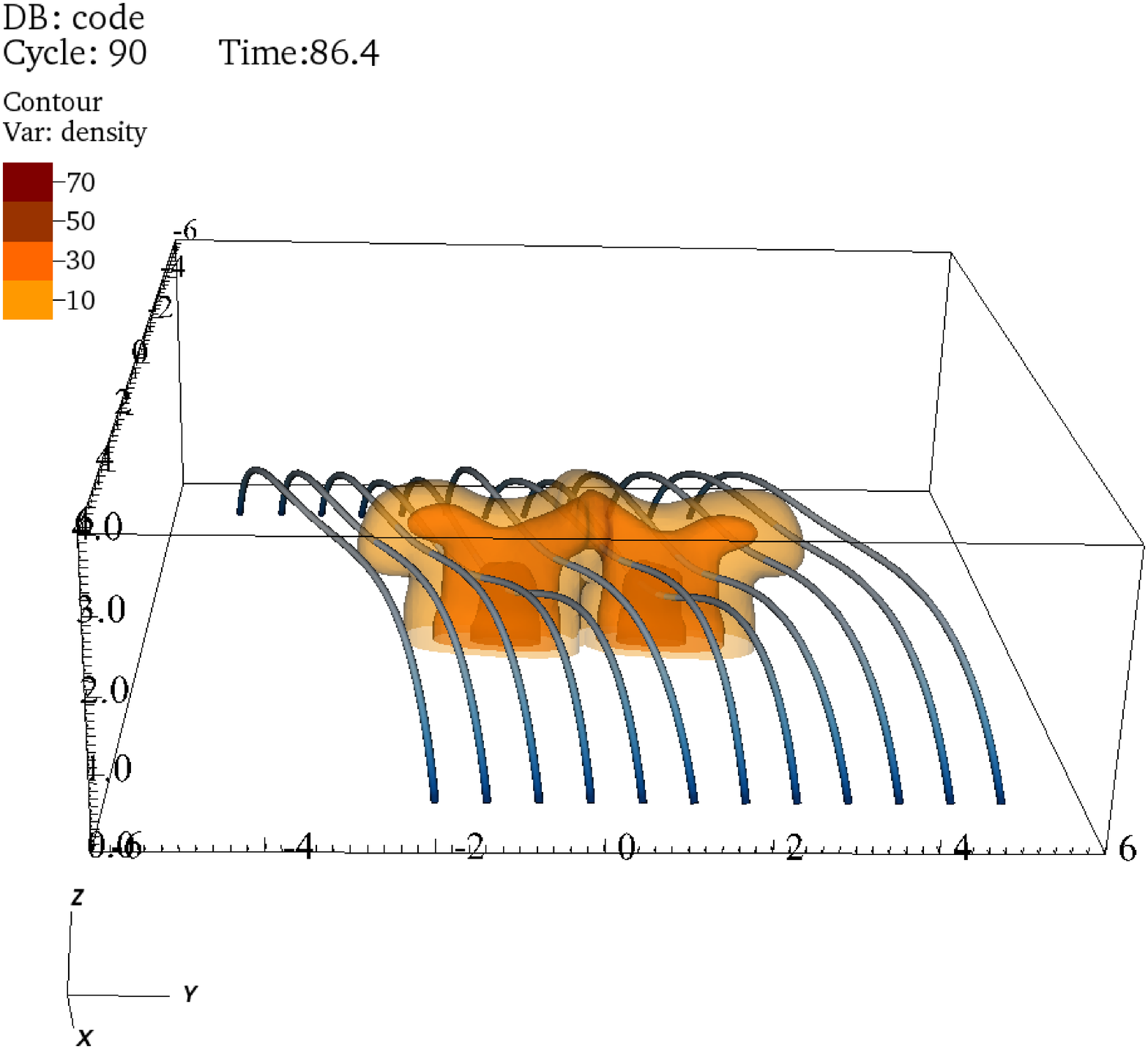}\\
\includegraphics[width=6.5cm]{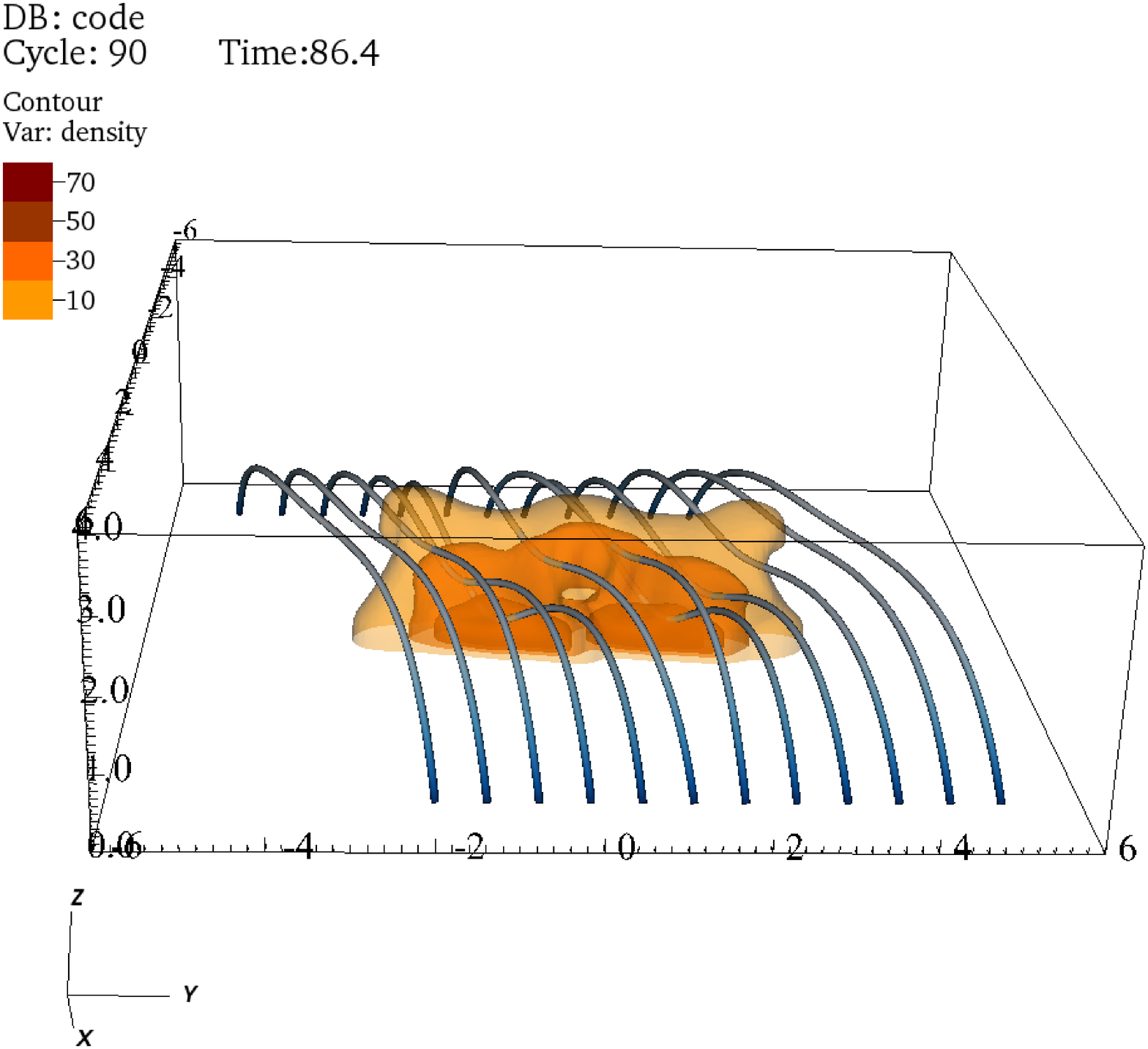} } \caption{\small Prominence density and
magnetic field for three different values of the magnetic field ($v_{A0}=20\, c_{s0}$,
$v_{A0}=15\, c_{s0}$, and $v_{A0}=10\, c_{s0}$ ) which are associated to the values of
$\beta$ of 0.075, 0.15, and 0.4 at the core of the prominence. Time is the same for the
three simulations.}\label{betadep} \end{figure}

The case analyzed in Sect.~\ref{0run} corresponds to a particular set of
parameters. It is interesting to study how the evolution of the system depends,
for example, on the strength of the magnetic field.  In Fig.~\ref{betadep} the
morphology of the prominence at a fixed time is shown for three different values
of the plasma$-\beta$ associated to different magnetic field strengths at the
core of the prominence. In the top panel we find the previous situation where
the magnetic field is able to provide the support to the dense material against
gravity. This case  corresponds to a value of $\beta=0.075$ at the end of the
simulation. This prominence can be classified
as detached. Middle and bottom  panels show a rather different scenario, and
they are associated to $\beta=0.15$ and  $\beta=0.4$, respectively. In these
last two cases the magnetic field is unable to provide the restoring force to
hold the whole prominence suspended above the photosphere. Most of the dense
plasma falls down and essentially pushes down the magnetic field near the
photosphere (see the field lines crossing the center of the prominence body in
Fig.~\ref{betadep}). For $\beta=0.15$  some voids in density are appreciated at
the center resembling some hedgerow prominences reported in observations
\citep[see the classification of prominences
by][]{menzelevans1953,jones1958,zirin1998}. For the case $\beta=0.4$ the
structure seems more compact resembling mound or curtain prominences with little
structure. Note that if the view of the prominence is along the $y-$direction
then it can resemble a pillar prominence which is simply a curtain prominence
seen edge-on.

\begin{figure}[!hh] \center{\includegraphics[width=6.5cm]{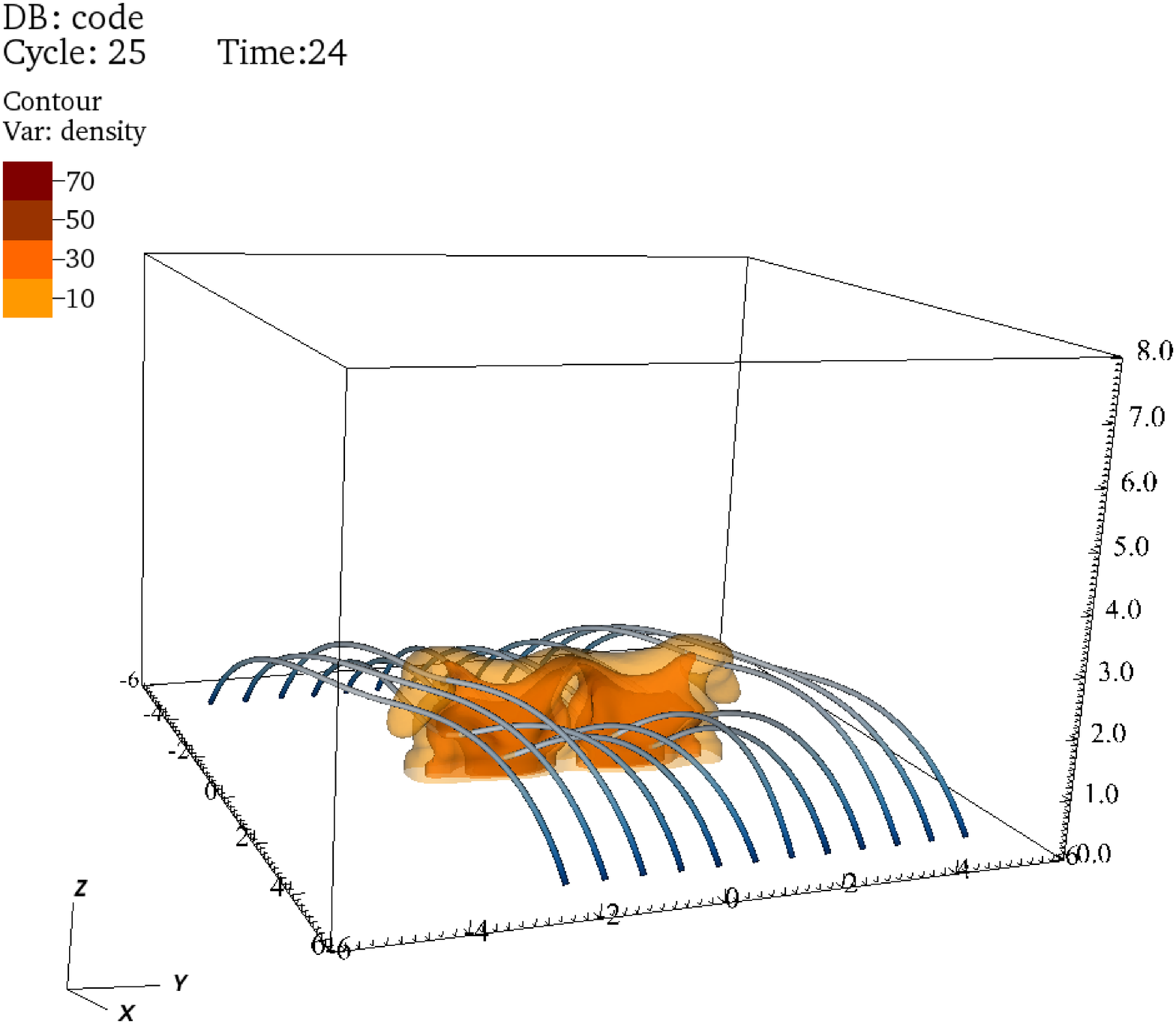}
\includegraphics[width=6.5cm]{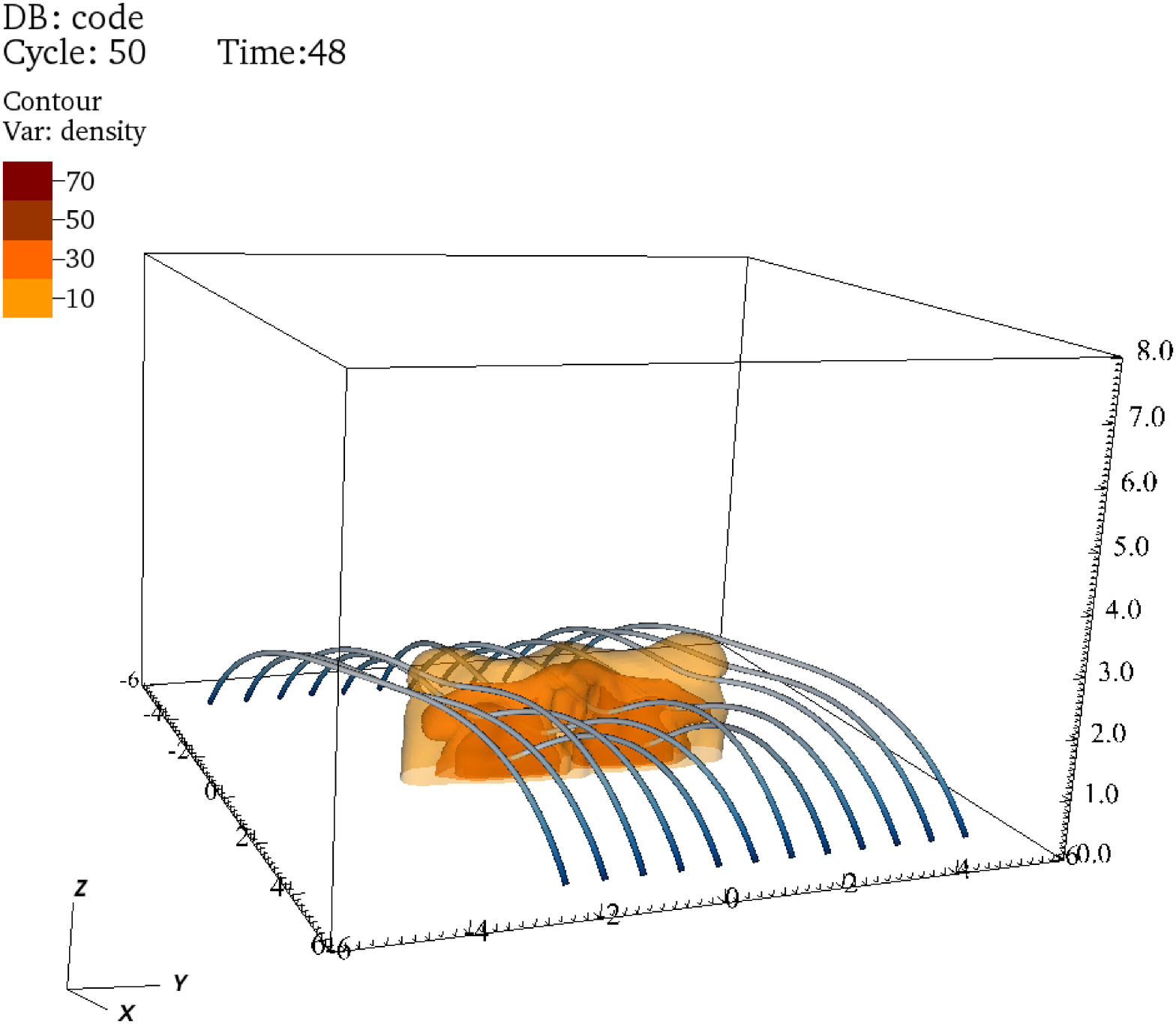}\\
\includegraphics[width=6.5cm]{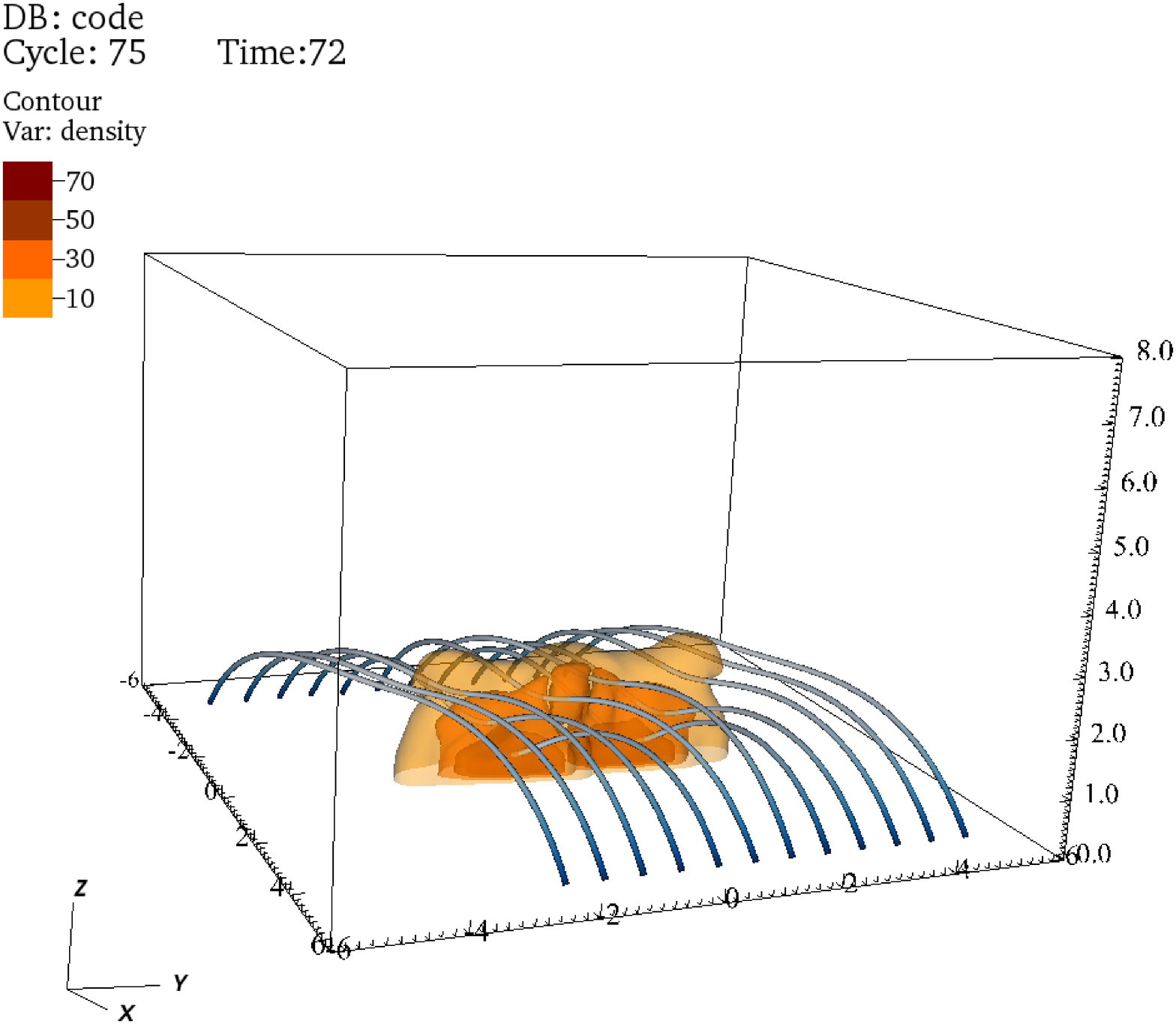} 
} \caption{\small Details of the
evolution of the prominence density for the
case $v_{A0}=10\, c_{s0}$. See also Movie3 in the online
material. }\label{densevol} \end{figure}

\begin{figure}[!hh] \center{ \includegraphics[width=6.5cm]{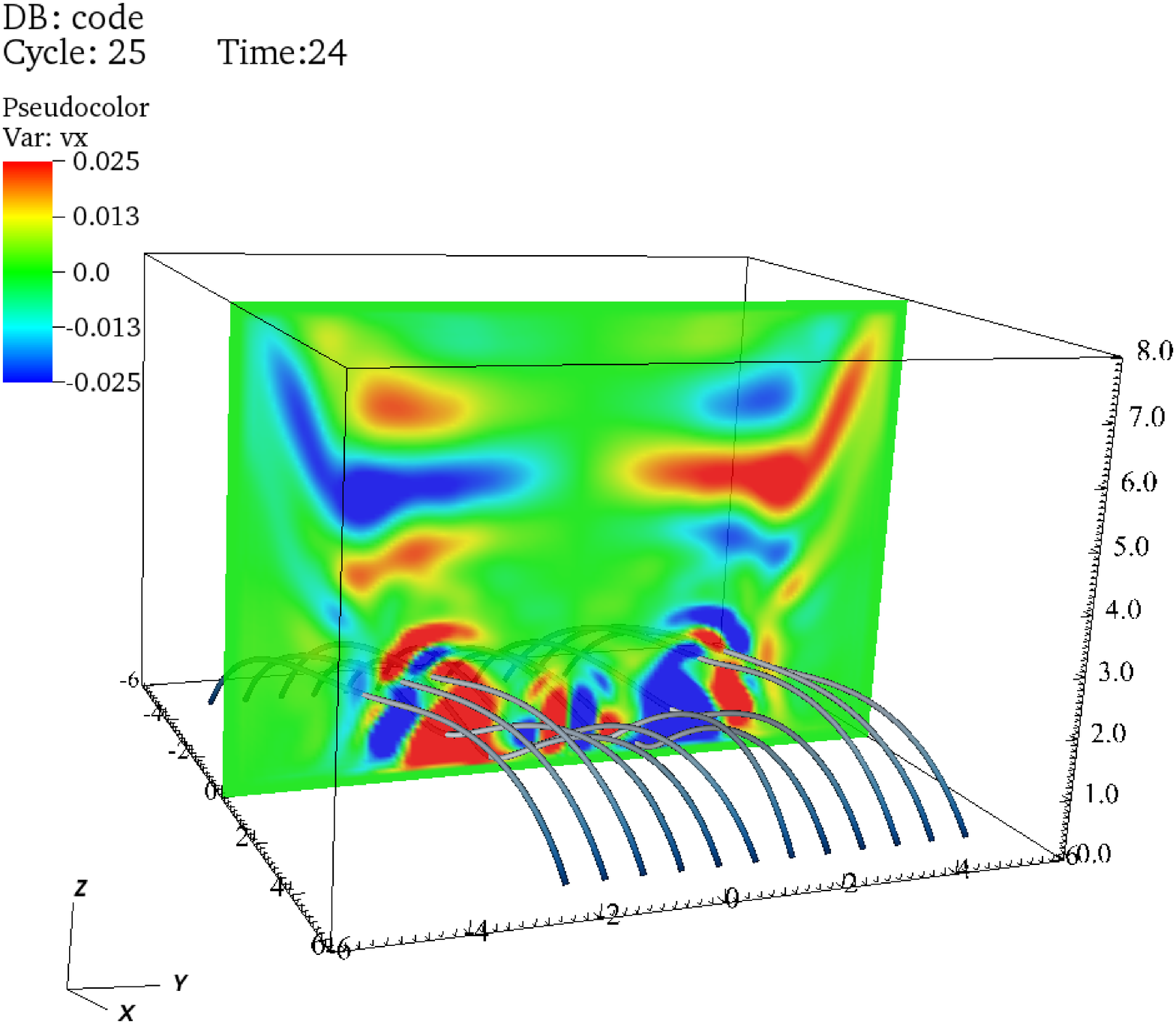}\\
\includegraphics[width=6.5cm]{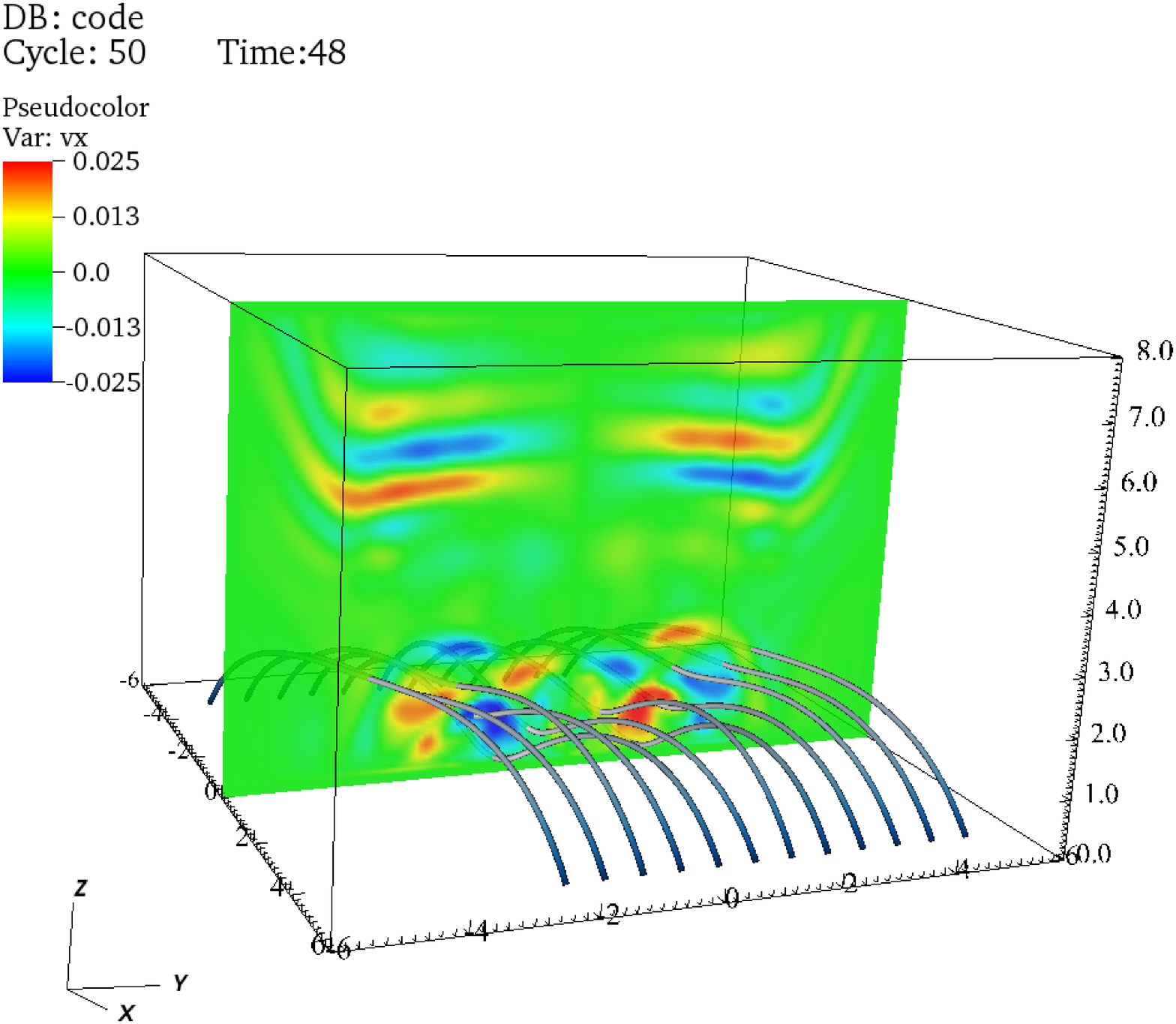} \\
\includegraphics[width=6.5cm]{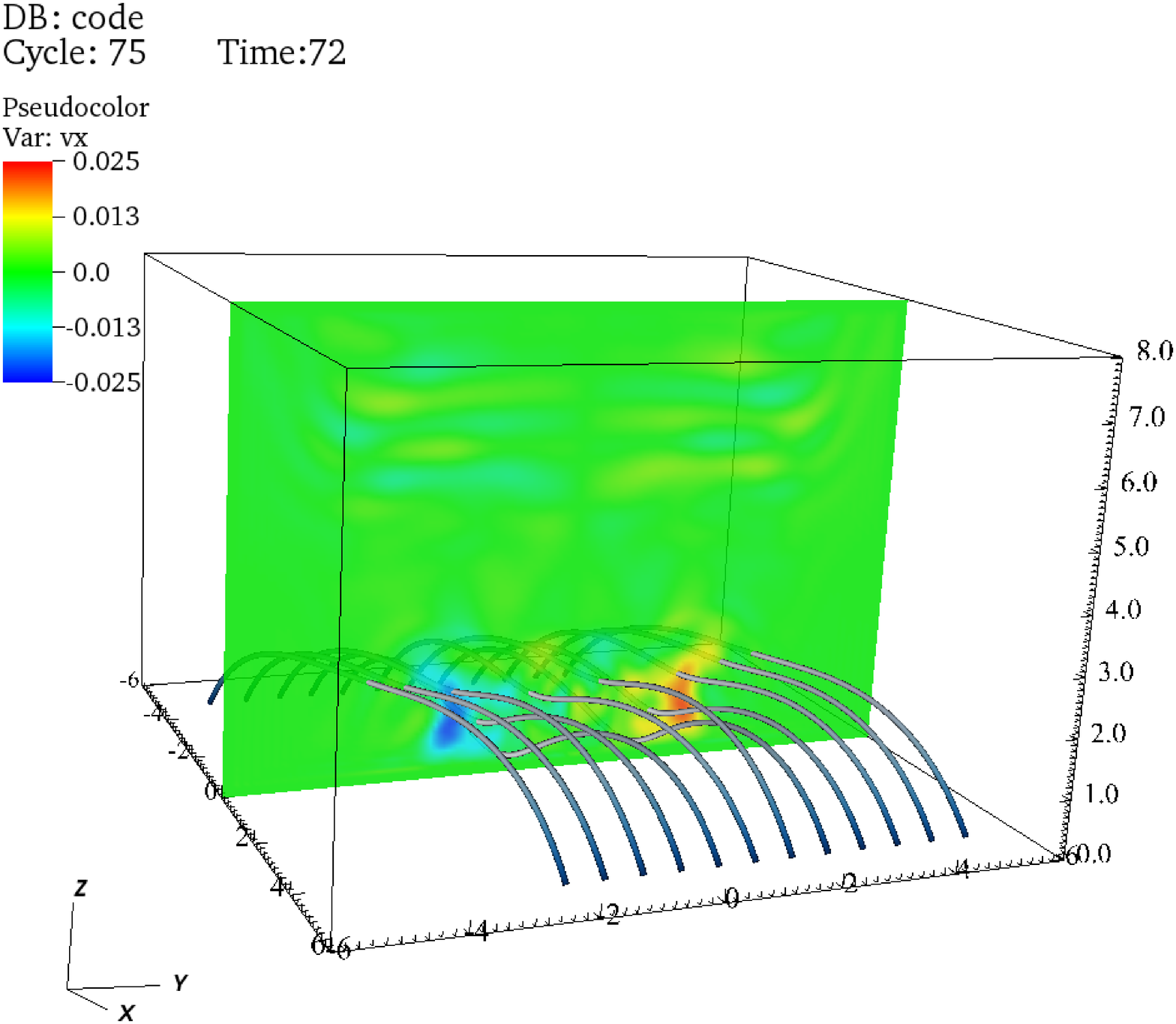} } \caption{\small Details of the
evolution of the horizontal component of the velocity ($v_x$) for the case
$v_{A0}=10\, c_{s0}$ and the same simulation as in
Fig.~\ref{densevol}.}\label{densevol1} \end{figure}

Details of the evolution for the curtain-like prominence are found in
Figs.~\ref{densevol} and ~\ref{densevol1} (see also Movie3). In these figures we
have represented the evolution of density and the $x-$component of the velocity,
$v_x$, passing through the central plane. The other velocity components are also
important at the beginning of the simulation but at the end of the simulation
$v_x$ dominates. At early stages of the evolution the behavior is quite
involved, as can be appreciated at the top panel of Fig.~\ref{densevol1},
velocities up to $4\, \rm km\, s^{-1}$ are found at the edges but also at the
core of the prominence. Outside the prominence we identify the emission of MHD
waves, this is specially clear near the top of the domain where several
wavefronts are visible. As time progresses the amplitudes of the motions
decrease both at the prominence and in the corona. This decrease in the velocity
amplitudes is mostly due to the energy leakage that takes place at the upper and
lateral boundaries of the computational domain. Numerical dissipation, although
small, also contributes to the attenuation of the motions. At later stages of
the evolution, see for example bottom panel of Fig.~\ref{densevol1}, we can
still find some velocity signals at the prominence which are related, in this
particular example, to a slow change in orientation of the axis of the
prominence since the motions in the $x-$direction alternate sign with respect to
the center of the structure.

\subsection{Dependence on prominence mass}

\begin{figure}[!hh] \center{\includegraphics[width=6.5cm]{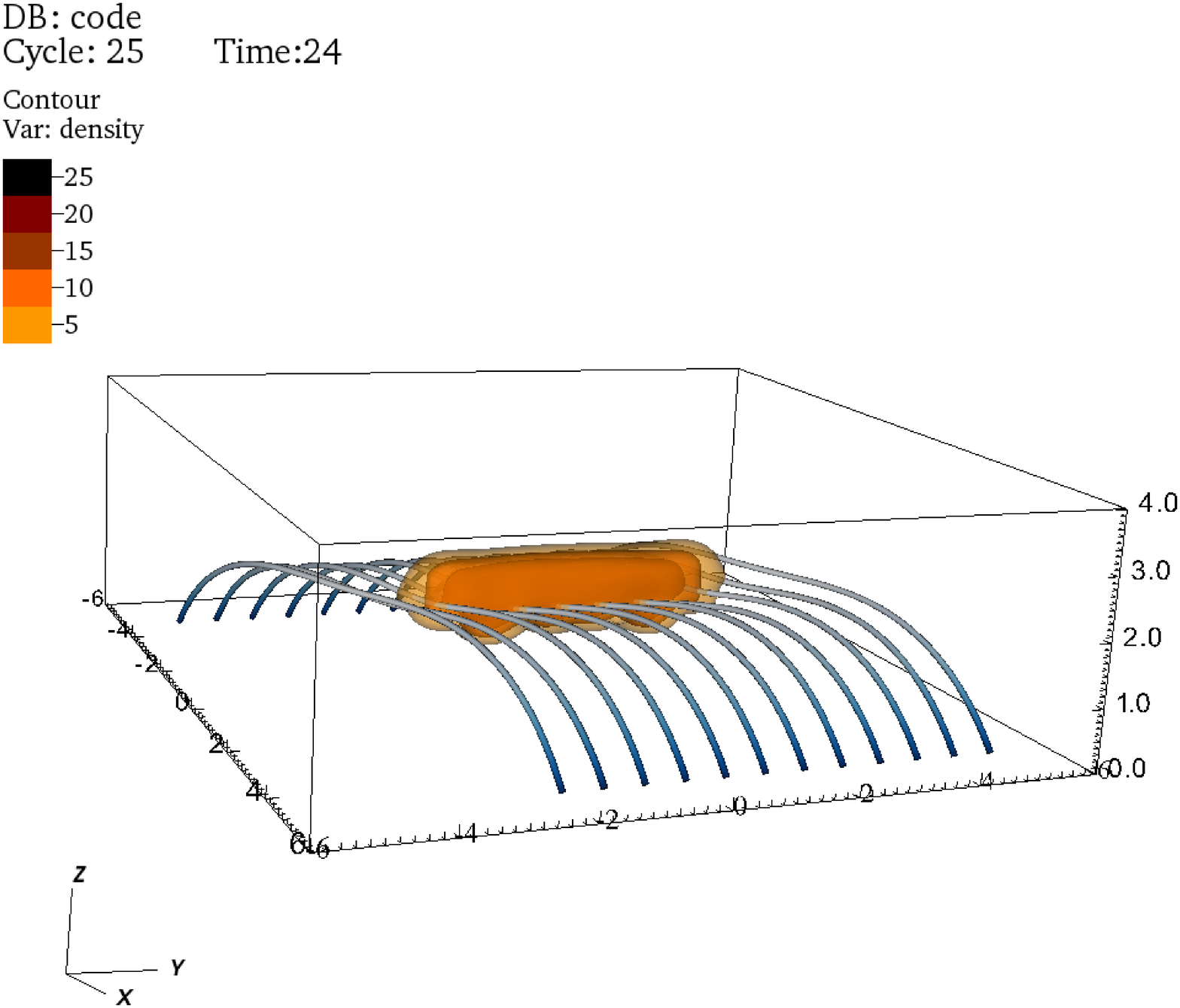}
\includegraphics[width=6.5cm]{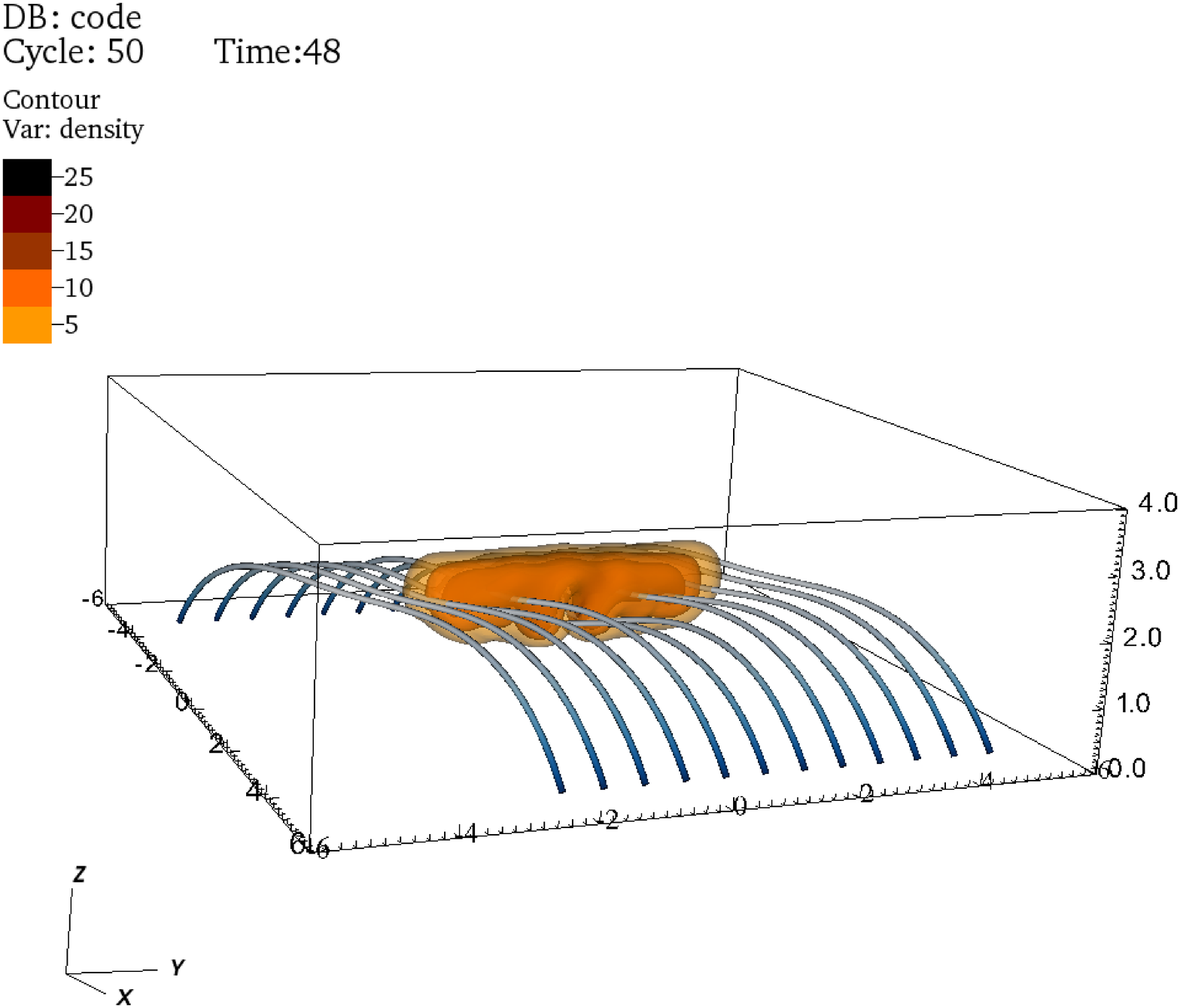}\\
\includegraphics[width=6.5cm]{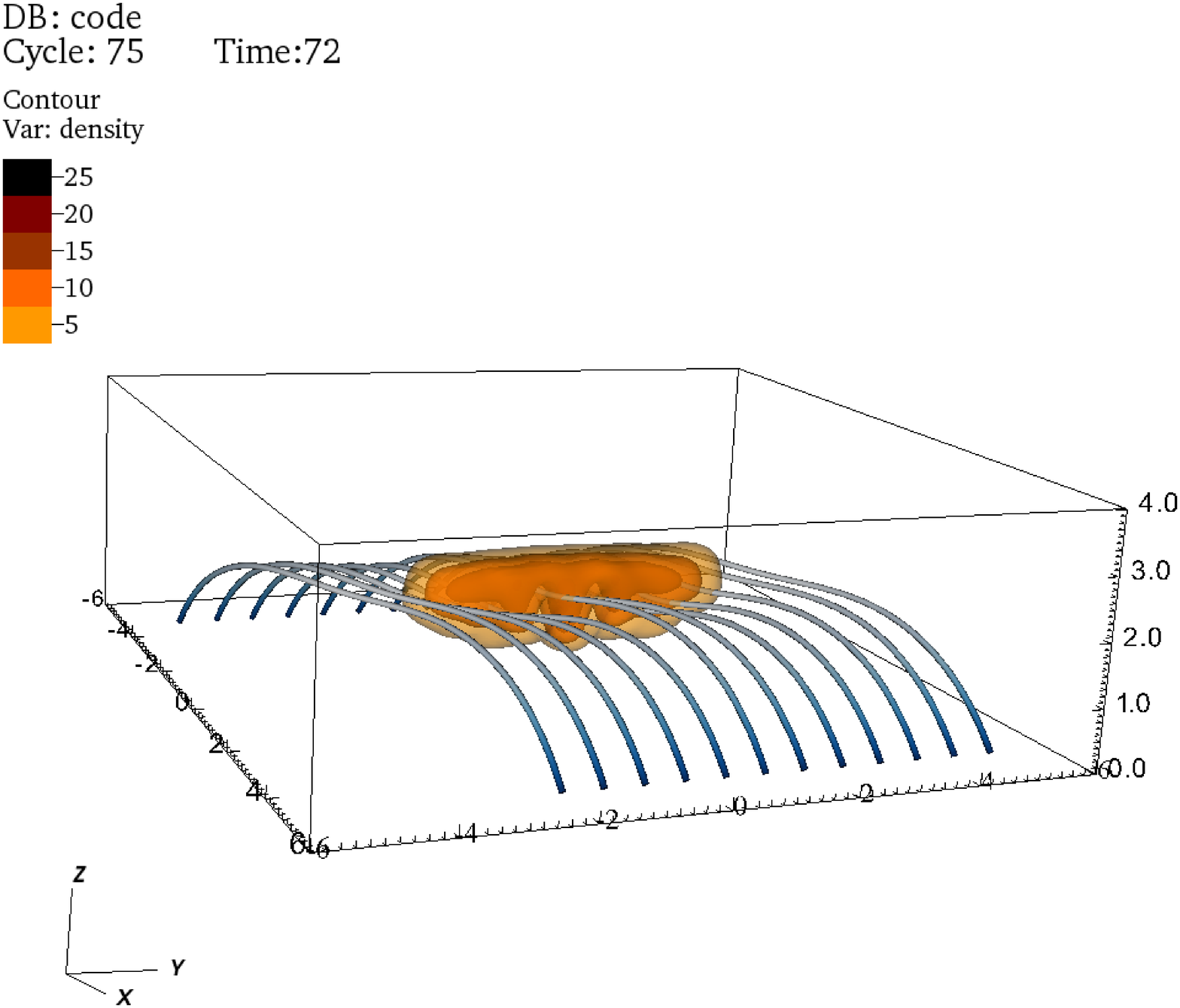}  } \caption{\small Details of the
evolution of the prominence density for the case $v_{A0}=10\, c_{s0}$ but for a total
prominence mass that is four times smaller than in the previous cases. The density
contrast between the prominence and corona is only 30 in this
simulation.}\label{densevol_mass} \end{figure}

Another relevant parameter is the total mass of the prominence. We have imposed a fixed
strength of the magnetic field at the base of the arcade, $v_{A0}=10\,c_{s0}$, and have
decreased the prominence mass with respect to the mound prominence studied in
Figs.~\ref{densevol} and ~\ref{densevol1}. Now the total mass is four times lower, i.e., of
the order of $3.25\times 10^{10}\, \rm kg$ which is still in the range of masses calculated
from observations (typically between $10^{9}-10^{12}\, \rm kg$). Three snapshots of the
density and magnetic field are plotted in Fig.~\ref{densevol_mass} for this case. Although
the plasma$-\beta$ at the core of the prominence is around 0.3, the magnetic structure is
able to sustain the prominence suspended above $z=0$, obtaining a detached structure. Thus,
the total mass loading of the prominence essentially determines, together with the
plasma$-\beta$, the morphology of the structure. Note also the appearance of the MRT
instability with some fingerprints visible in Fig.~\ref{densevol_mass}.

\subsection{Dependence on magnetic shear}\label{sheardepsec}

\begin{figure}[!hh] \center{\includegraphics[width=6.5cm]{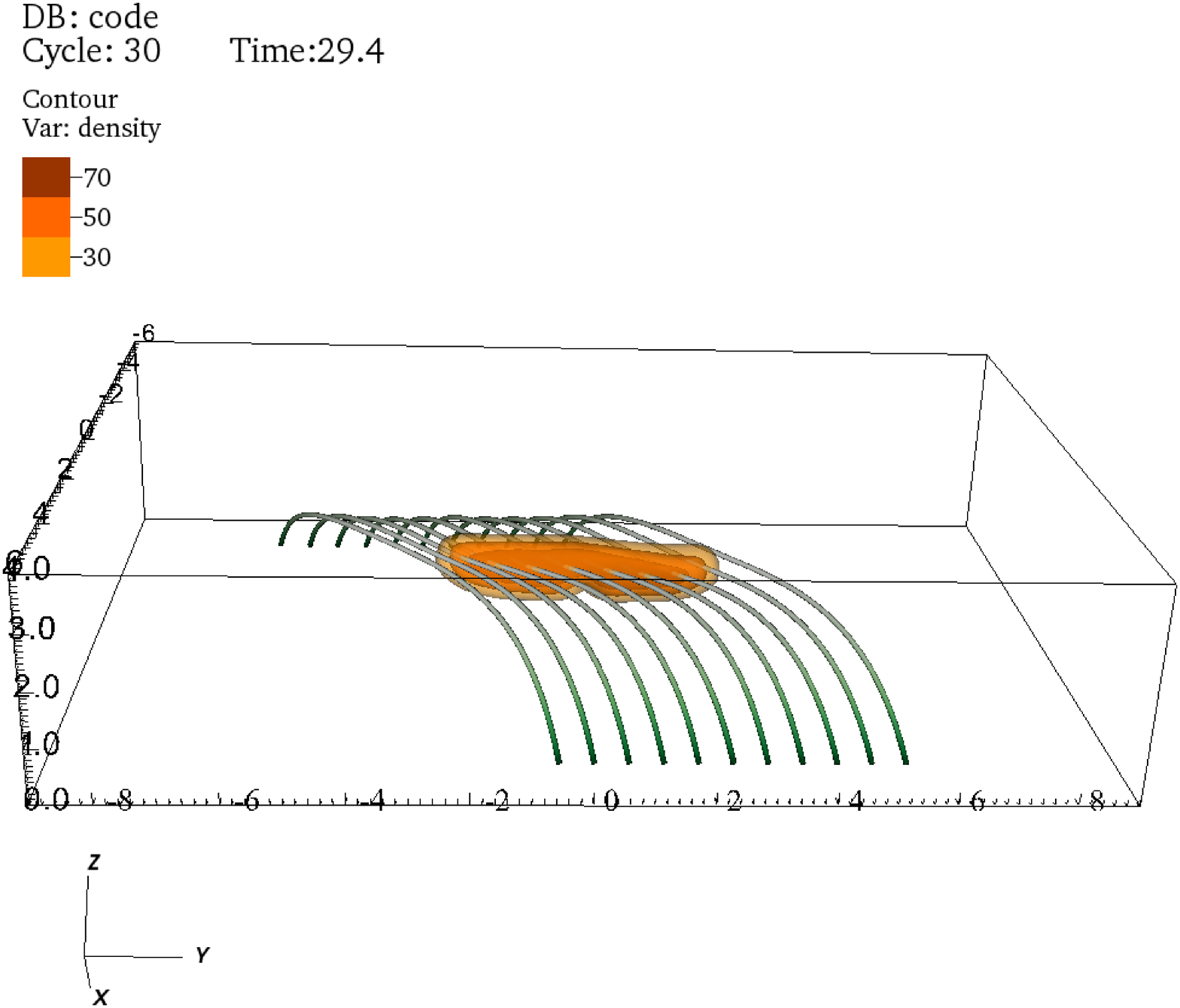}
\includegraphics[width=6.5cm]{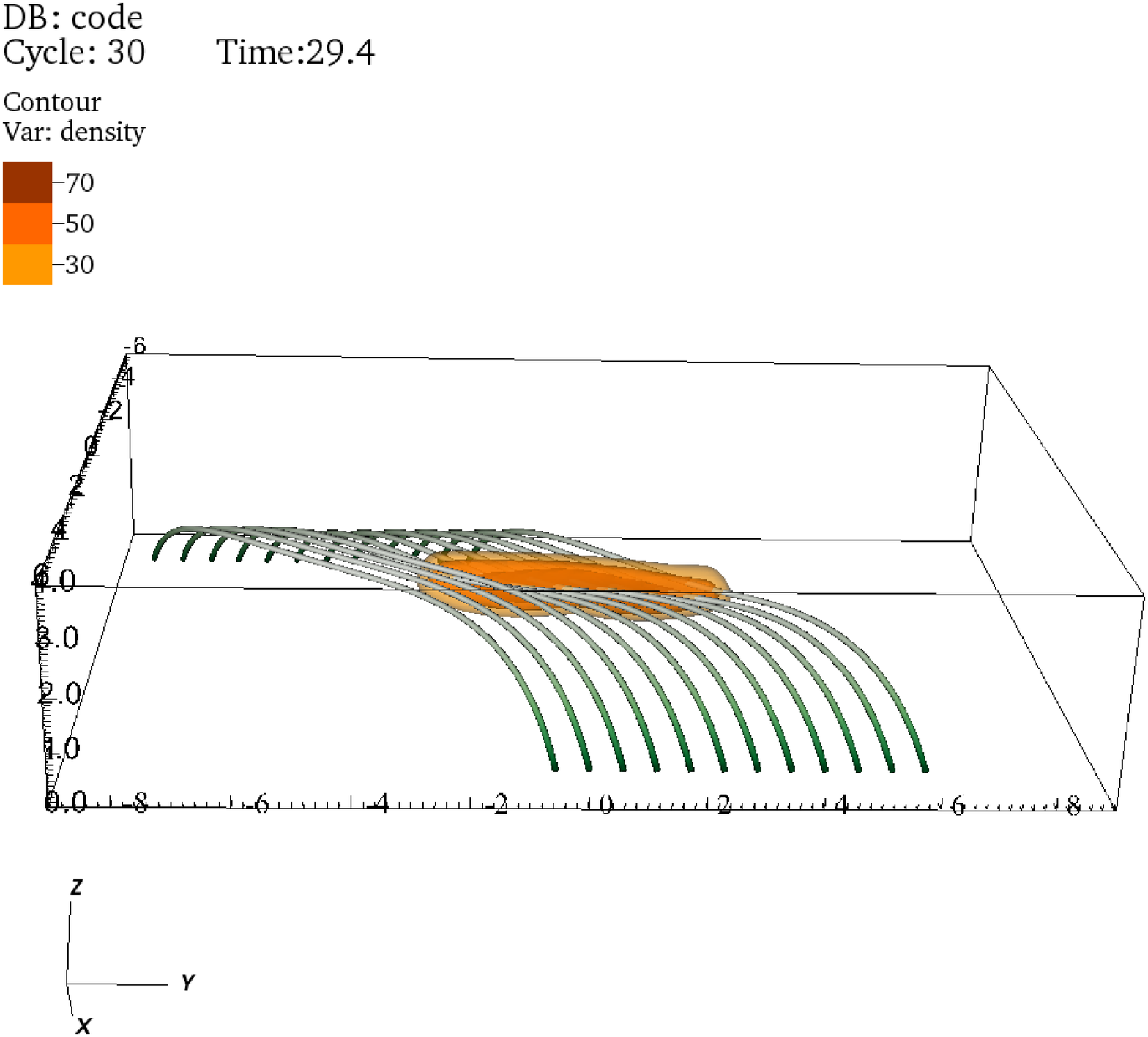} \includegraphics[width=6.5cm]{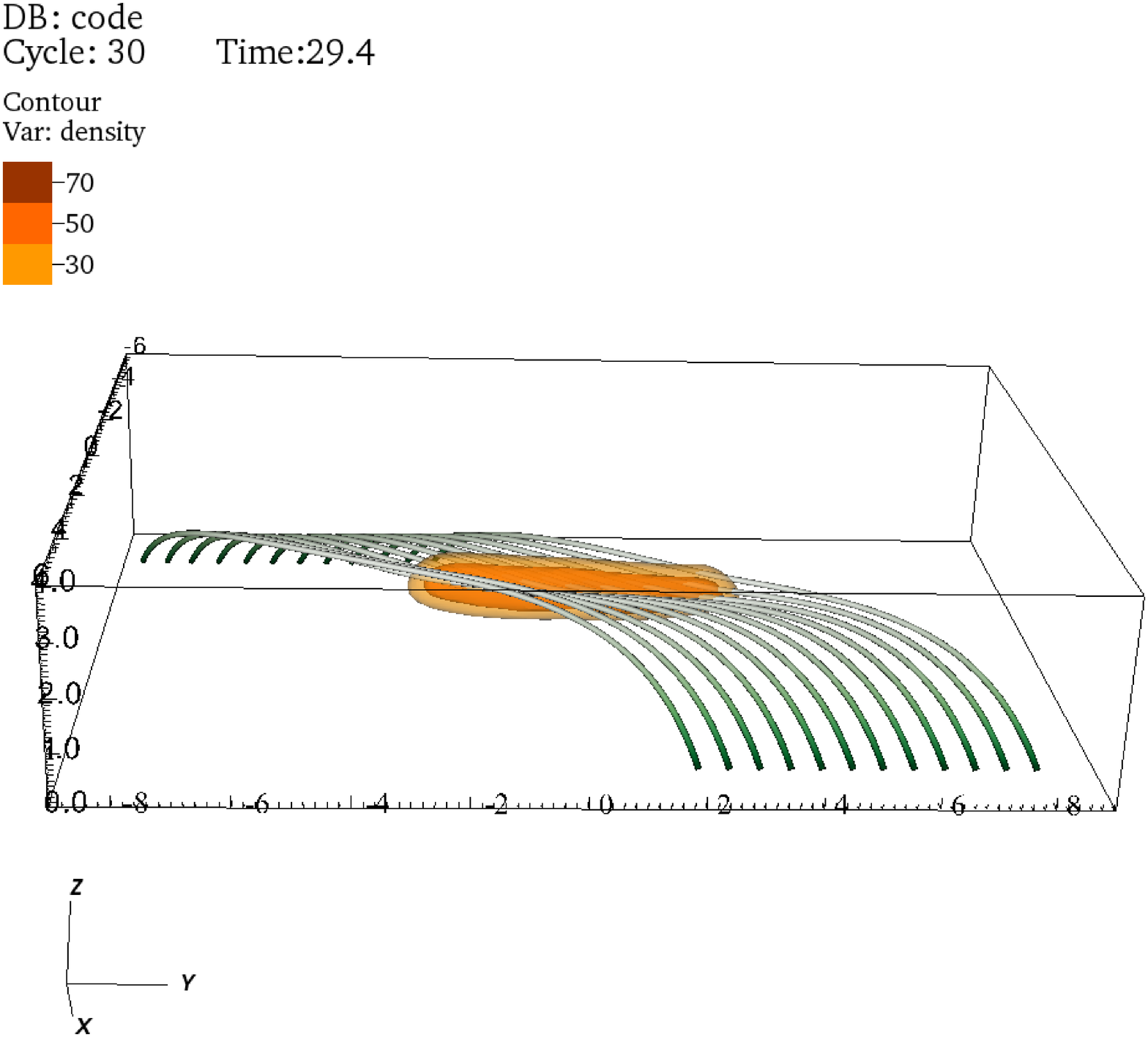} }
\caption{\small Density and magnetic field distribution for three different values of the
magnetic shear, $l/k=0.85$, $l/k=0.75$, $l/k=0.65$, that correspond to typical shear angles
of $32^{\circ}$, $41^{\circ}$, and $54^{\circ}$, respectively. The width of the domain in
the $y-$direction is larger than in the previous simulations since an increased shear leads
to longer field lines.}\label{sheardep} \end{figure}

\begin{figure}[!hh] \center{\includegraphics[width=8cm]{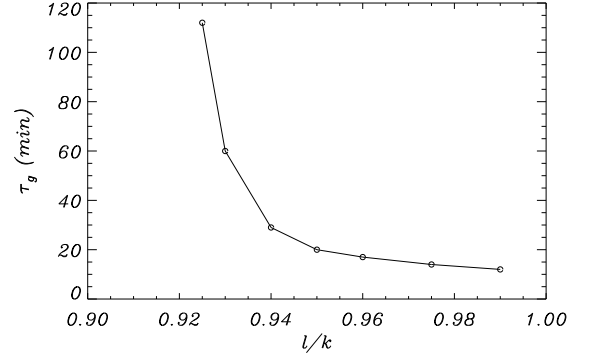}} \caption{\small
Growth-time of the dominant unstable MRT mode as a function of $l/k$. The case $l/k=1$
corresponds to the situation without shear. For $l/k<0.9$ the magnetic structure is
unable to support the prominence and the material falls to the
photosphere.}\label{taulok} \end{figure}

A parameter that can be also changed in the configuration is the amount of shear
in the arcade. We have performed a set of different simulations by changing this
parameter and the results are found in Fig.~\ref{sheardep}. For example, the MRT
instability already developed for the case $l/k=0.95$ at $t=29.1\,\rm min$ (see
Fig.~\ref{promva20side_1} middle panel) is not present for the arcade with
$l/k=0.85$. For larger values of the shear the global shape of the prominence
does not change much from the side view of Fig.~\ref{sheardep}. This means that
shear helps to have a more stable configuration regarding the MRT instability.
In order to quantify this effect we have plotted in Fig.~\ref{taulok} the
estimated growth-rates for different values of  $l/k$. The unsheared arcade, the
case $l/k=1$, has the shortest growth-time which is around $20\,\rm min$. The
curve shown in Fig.~\ref{taulok} indicates that the growth-time increases with
shear, meaning that this parameter helps to reduce the instability. Note that
the dependence of the growth-times with shear can be quite strong, and that a
small amount of shear may significantly reduce the instability. For example, we
see that for  values of $l/k$ below 0.92 the growth-times are larger than 2
hours. For this equilibrium parameter the arcade is still weakly sheared since
it would be an intermediate case between the cases $l/k=0.95$, see
Fig.~\ref{promva20side_1}, and $l/k=0.85$, represented in Fig.~\ref{sheardep}
top panel.

\begin{figure}[!hh] \center{\includegraphics[width=8cm]{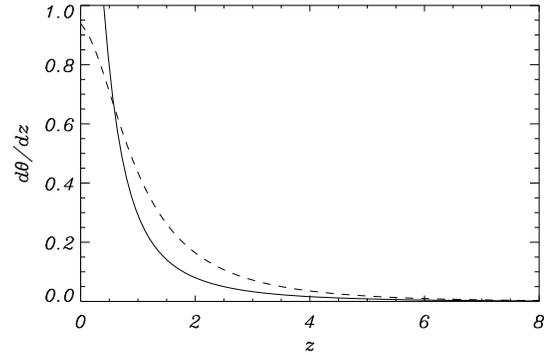}}
\caption{\small  Derivative of the shear angle (defined as
$\theta=\arctan(B_y/B_x)$) respect to $z$ at the center of the system ($x=y=0$)
as a function of height. The continuous line corresponds to the case $l/k=0.95$
while the dashed line represents the case $l/k=0.65$.}\label{derivshear}
\end{figure}

We have investigated the reasons of the stabilization effect produced by an increase
in magnetic shear. We have found basically two reasons. The first one is related to the
strength of the magnetic field at the prominence body. It turns out that a decrease in the
parameter $l/k$ (increase in shear) produces an increase in the horizontal component of
the magnetic field at the prominence. For example, for $l/k=0.65$ the horizontal component
is 1.4 times larger than for $l/k=0.95$. Stronger horizontal magnetic fields at the core
of the prominence lead to a larger magnetic tension, and this is the force that helps to
stabilize the MRT unstable modes. The second reason is related to the change of the shear
angle with height. The effect of shear on MRT instabilities has been recently investigated
analytically by \citet{rudermanterradas2014} in a much more simple geometry, a 2D slab
configuration (in the $xz-$plane) with a discontinuous change in the magnetic shear and
including propagation in the perpendicular direction (the $y-$direction). These authors
have found that the growth rate is bounded under the presence of magnetic shear. The
important result is that for small shear angles the growth time of the most unstable mode
is linearly proportional to the shear angle, while in the limit of large angles the growth
time is essentially independent of the shear angle. In our arcade configurations the
situation is more complex since the magnetic shear changes smoothly with height (see
Fig.~\ref{derivshear}). This must affect the growth times of the instability. Using the
results of the simple slab model we can explain the qualitative behavior in the growth
times with the parameter $l/k$. For $l/k$ large, the variation of shear with height is
relatively small (see in Fig.~\ref{derivshear} the region around the prominence location,
i.e., around $z=2$), meaning that the effective shear is small. Under such conditions the
growth time is small (fast development of the instability). On the contrary, when $l/k$ is
small the change of shear with height is stronger (see Fig.~\ref{derivshear}) meaning that
the effective shear is larger and thus the growth time is longer. This provides a
qualitative agreement with the results shown in Fig.~\ref{taulok}.

\begin{figure}[!hh] \center{\includegraphics[width=8cm]{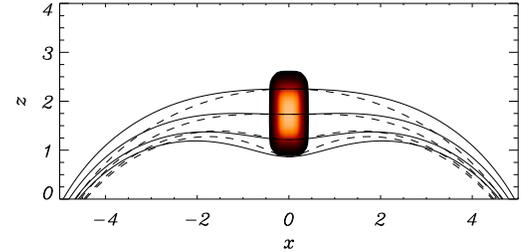}}
\caption{\small Selected magnetic field lines for the case $l/k=0.95$ (continuous lines)
and $l/k=0.65$ (dashed lines). Note the difference in the depth and width of the
magnetic dips for the two configurations.}\label{dips} \end{figure}

In fact, an increase in shear changes the depth of the dips, i.e., the
vertical distance between the dip centre and the highest point on a magnetic
field line, and it also changes the width of the dips. In Fig.~\ref{dips} the
magnetic field lines (projected in the $xz-$plane) crossing the initial density
distribution at a given height are plotted for two different values of the
shear. For the case $l/k=0.95$  (continuous lines)  we see that the field lines
are quite flat at the top of the prominence and the depth of the dip increases
as we move toward the photosphere. For the configuration with $l/k=0.65$ (dashed
lines) with a much stronger shear the situation is different, at the top of the
prominence the field lines are more curved than for $l/k=95$ and potentially
less favorable to support the dense material, but as we move to lower heights
the depth of the dips increases with respect to $l/k=95$ and thus helping to
sustain the mass. On the contrary, the width of the dips is smaller than for the
case $l/k=0.95$. Therefore, the magnetic support can be slightly different for
the two configurations.

Although in  Fig.~\ref{sheardep}  (bottom panel) the prominence body seems to keep a
compact structure, a view from the top of the configuration (not shown here), reveals a
rather different behavior. The dense material tends to move along the magnetic field, and
strong deformations appear at the edges of the prominence. In fact, this part of the
prominence eventually descends toward the photosphere. This is in agreement with the
results of \citet{karpenetal2001,lunaetal12a} that have found that in a highly sheared
arcade with shallow dips the condensation of cool material is short-lived and falls
rapidly to the chromosphere. Another effect that we find in the simulations is that in the
meantime the orientation of the main axis of the prominence changes with time and tends to
be aligned with the direction of the magnetic field. These features, that might be
relevant regarding observations, need further investigation but are left for future
studies since a detailed analysis would require much longer total simulation times with
better spatial resolutions.

It must be pointed out that in the analysis of the dependence of the instability
with shear ($l/k$) we have imposed that $B_2=B_1$ in Eqs.~(\ref{bx})-(\ref{bz}).
This condition selects a particular family of magnetic arcade solutions. Other
choices of the constants $B_1$ and $B_2$ might lead to the presence of
significant dips even for cases where shear is strong. Hence, it might be
possible to find configurations with strong shear in which the mass of the
prominence does not fall to the photosphere.

\subsection{An irregular prominence}

\begin{figure}[!hh] \center{\includegraphics[width=6.5cm]{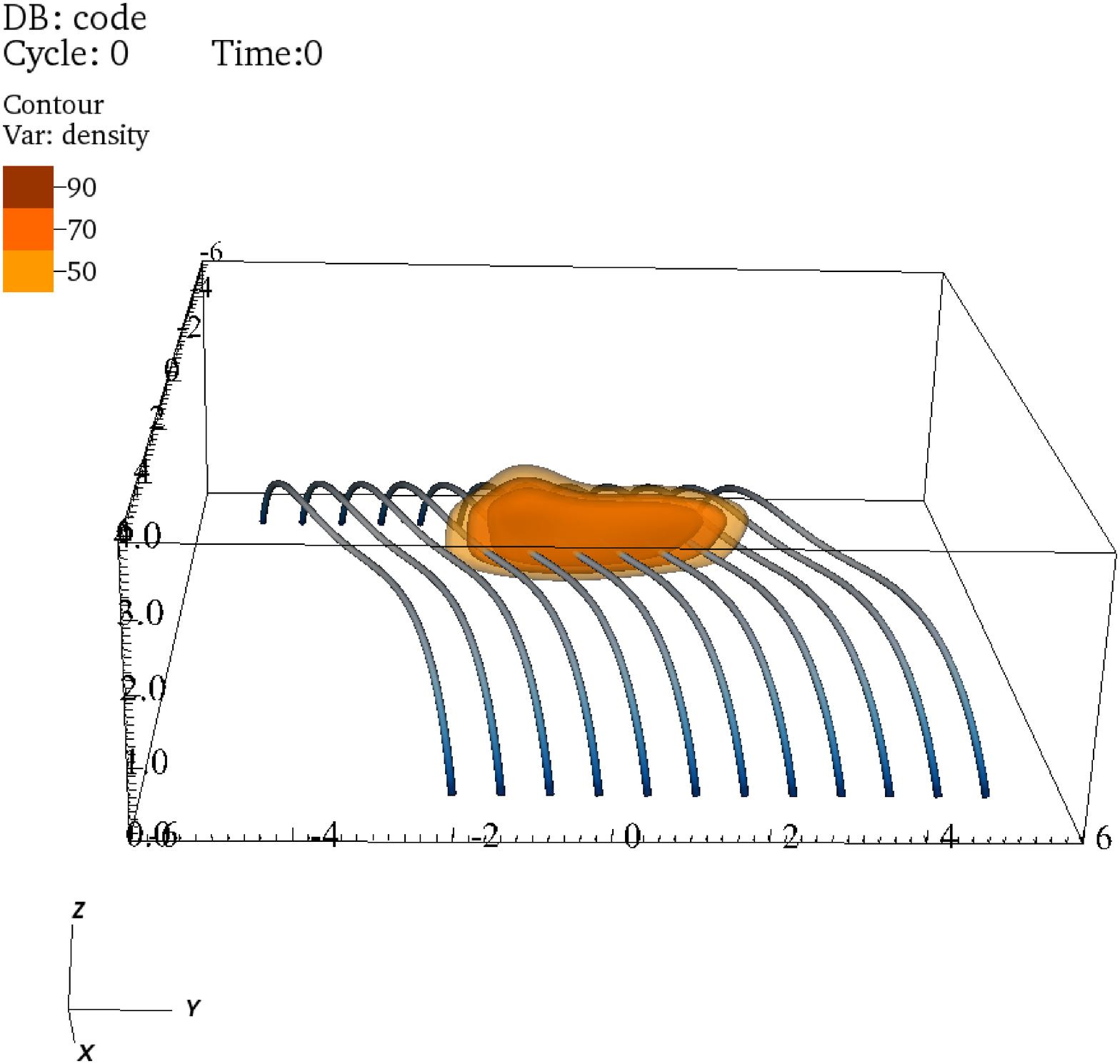}
\includegraphics[width=6.5cm]{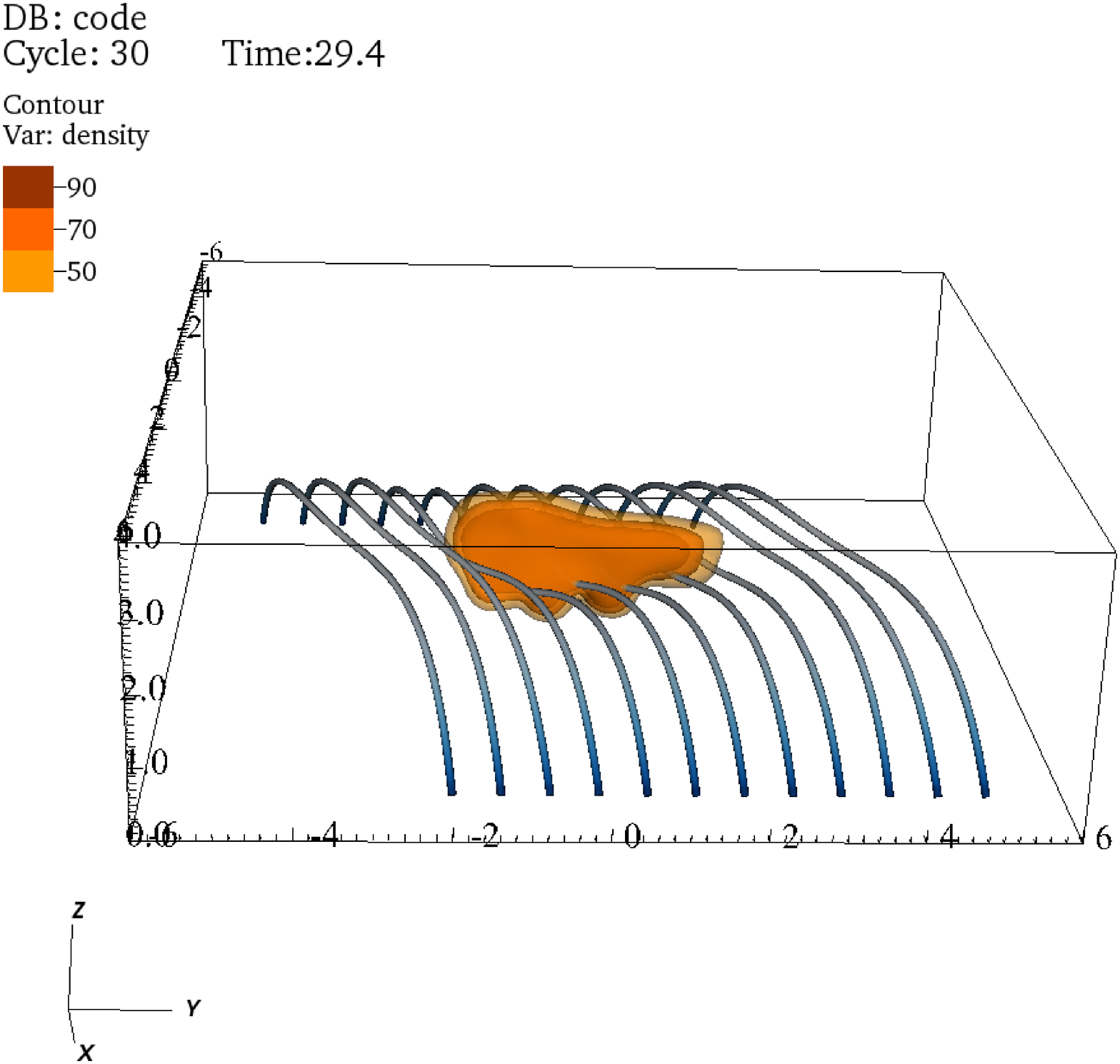}
\includegraphics[width=6.5cm]{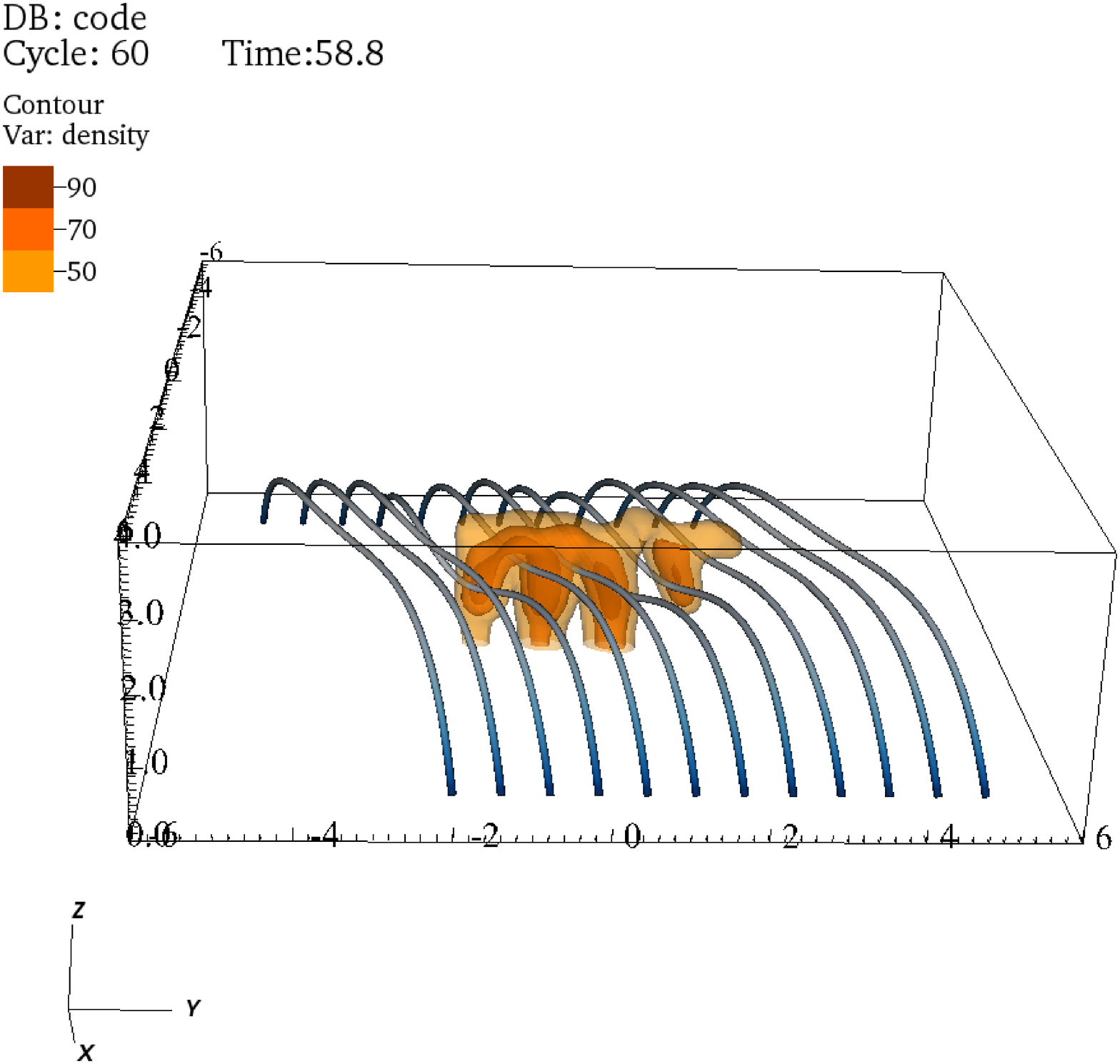}} \caption{\small Density
isocontours at different times. Fingers and plumes associated to the MRT
instability are present in this simulation. The difference with respect to
Fig.~\ref{promva20side_1} is in the initial shape of the density distribution
which is now non-symmetric, the rest of parameters are the
same.}\label{promva20dens_1} \end{figure}

Finally, it is important to remark that all the simulations presented in the
previous sections are based on the assumption that the initial mass deposition
is symmetric. The results are thus affected by this assumption, and among other
things, only symmetric modes are excited in the system. Since observations show
that in general the distribution of mass in the prominence body is very
inhomogeneous it is also interesting to investigate a more general case, and see
the effect of inhomogeneity on the numerical experiments. For this reason, here
we present the results of a run with an irregular mass. The initial density
distribution together with the time evolution are represented in 
Fig.~\ref{promva20dens_1}. Most of the initial mass is concentrated around the
left part of the prominence, and the interface between the core and corona is
not purely horizontal. Again the prominence body is pushing the magnetic field
in the negative vertical direction due to the effect of the gravity force. At
later times, around  $t=30\,\rm min$, we identify the initial stages of the
development of the MRT instability at the curved interface at the bottom of  the
prominence. Plumes and fingers are again quite evident at $t=59\,\rm min$. The
development of the instability proceeds at a faster rate than in the purely
symmetric prominence. Interestingly, these fingers may eventually connect to the
photosphere, meaning that the morphology of the structure might evolve from a
detached prominence to a hedgerow prominence.

\begin{figure}[!hh] \center{\includegraphics[width=6.5cm]{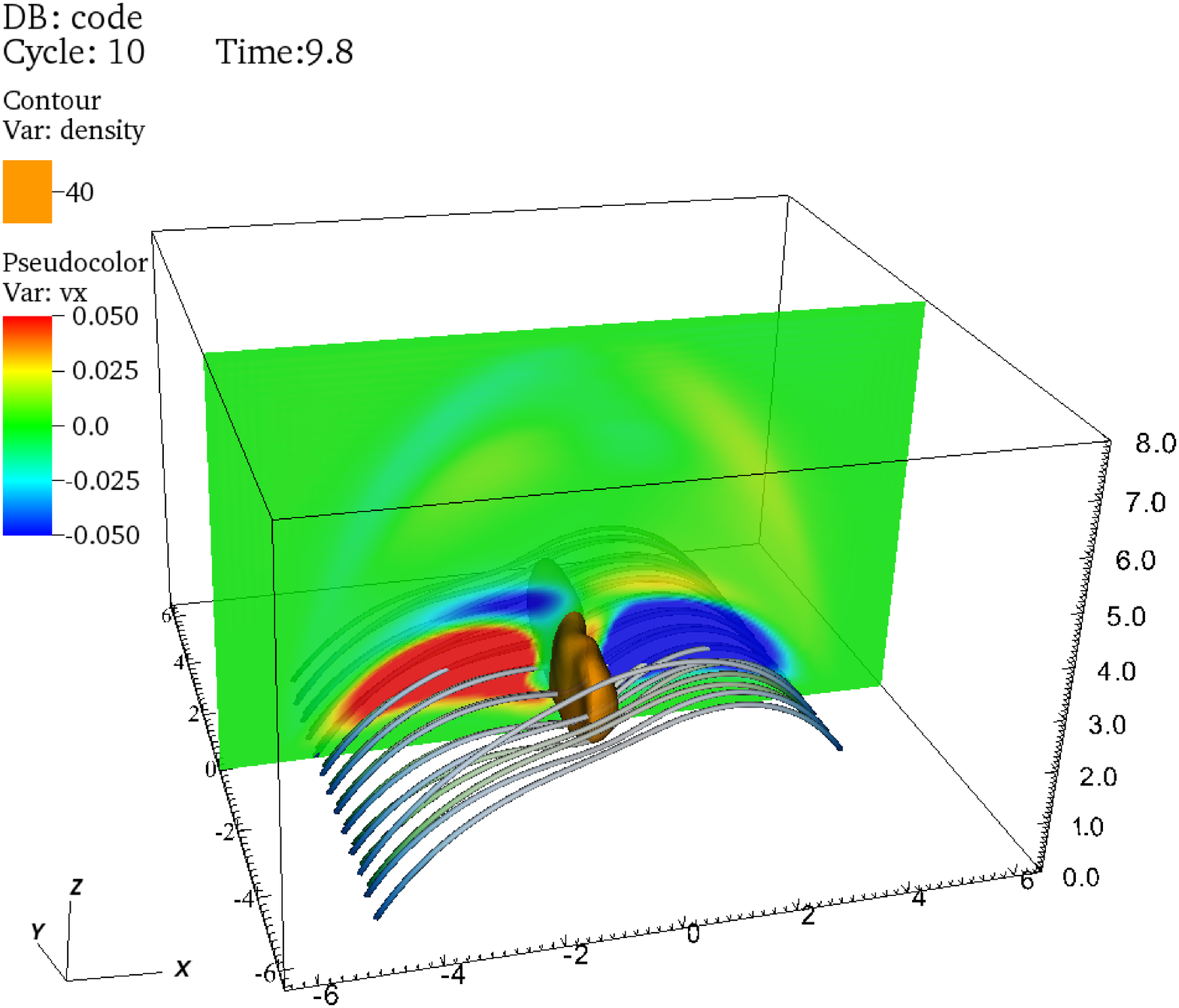}
\includegraphics[width=6.5cm]{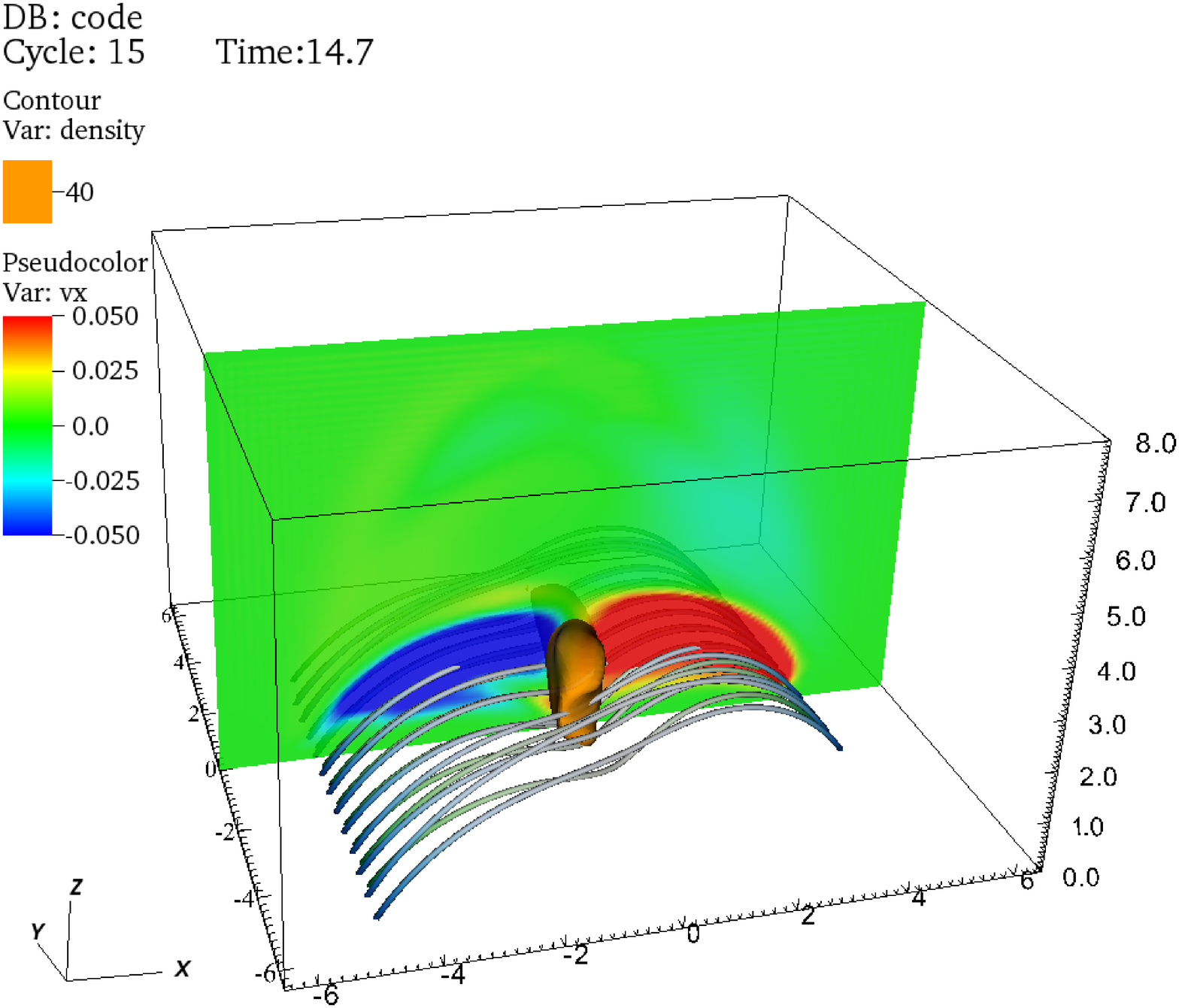}
\includegraphics[width=6.5cm]{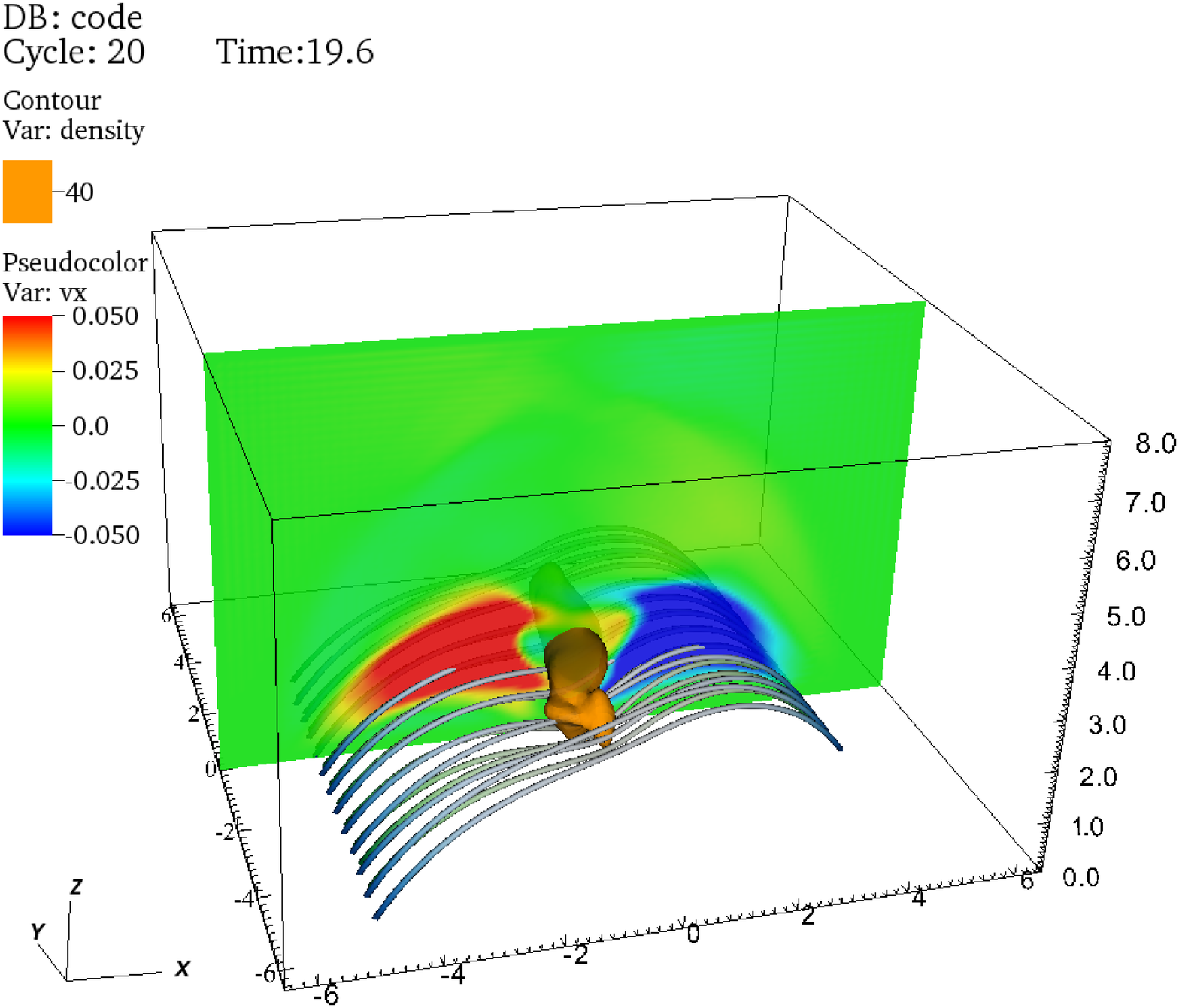}} \caption{\small Time evolution of 
density and horizontal  velocity at the $xz-$plane at $y=0$ at the initial stages of the
evolution and before the onset of the MRT instability. See Movie4 in the online
material.}\label{promva20velinh} \end{figure}

In this simulation since the system is asymmetric the excitation of other types
of motions with respect to the symmetric case are possible. For example, in
Fig.~\ref{promva20velinh} the velocity component in the $x-$direction together
with the density are represented as a function of time. Now the point of view
has been changed in the plot to better visualize the results. In this plot we
identify motions in $v_x$ that alternate sign with respect to the location of
the prominence body around the center of the system. These motions are
associated to the excitation of perturbations running mostly  parallel to the
magnetic field and therefore linked to the excitation of slow magnetoacoustic
waves \citep[see][for the equivalent results in 2D]{terradasetal2013}. As a
consequence of these motions the whole prominence shows changes in the
cross-section, as can be appreciated in Fig.~\ref{promva20velinh} (see density
isocontours), and at the same time  it is oscillating laterally, mostly along
the $x-$direction (see Movie4). This can have close links with large amplitude
longitudinal oscillations reported in the literature
\citep[see][]{tripatetal09,zhangetal12,lunaetal2014}.

\section{Discussion and conclusions}

We have investigated the time evolution of a simple three-dimensional density
enhancement representing a prominence embedded in a sheared coronal arcade. We
have studied the evolution of the prominence mass and the magnetic field in a
selfconsistent way. This is different respect to other studies about the
topology of the magnetic field inferred from photospheric extrapolations since 
the prominence mass is not taken into account in the model. The results from our
study indicate that in general the dynamics of the coupled prominence mass to
the magnetic field is quite involved. We have tried to understand how the
results depend on the different parameters of the configuration. A critical
parameter that determines the characteristics of the evolution is the
plasma$-\beta$. For low$-\beta$ configurations cold material can be suspended
above the photosphere, and it can represent a detached prominences typically
reported in observations. In this case there is a balance between the prominence
weight, that depends on the mass, and the Lorentz force that changes with the
plasma-$\beta$.  A rather different configuration is found when $\beta$ is
increased. In this last case we find that the dense material is essentially
moving toward the photosphere and compressing the magnetic field. The overall
shape of the structure is more similar to hedgerow or even curtain prominences.
Hence, the global morphology of the prominence is highly determined by the
plasma$-\beta$ but it is also affected by the boundary conditions at the
photosphere. Here, we have focused on  line-tying conditions and have found that
they are crucial for magnetic support since they communicate the effect of the
dense photosphere on the prominence body. The total mass loading of the
prominence is also another parameter that determines the morphology and
evolution of the structure.


For prominences supported against gravity in the low$-\beta$ regime, we have
found fingerprints of the excitation of MRT unstable modes. Fingers and plumes
develop on a global scale affecting the whole prominence body. Although the
nature of the flows found in our numerical experiments are due to the release of
gravitational potential energy, the flow pattern is more complex than in the
standard MRT instability at an interface that it is in equilibrium. The reason
is that during the development of the instability the prominence is oscillating
as a consequence of line-tying conditions and this affects the growth-rates of
the modes. The spatial scale of plumes and fingers found in the simulations are 
much larger than those commonly reported in observations by, for example, 
\citet{bergeretal10} and studied from the numerical point of view by
\citet{hillieretal11,hillieretal12}.  In our simulations we are simply not
resolving those spatial scales due to a lack of spatial resolution. The voids in
density found in our simulations can be associated to the appearance of large
scale cavities/arches or even bubbles commonly reported in observations, 
specially in polar crown prominences. Other interpretations have been proposed
to explain bubbles, for example \citet{ryuetal10} have suggested that they are
the consequence of the excitation of the kink unstable mode in a flux-rope
prominence configuration. \citet{dudiketal2012} have proposed that the
formation of cavities is due to the emergence of magnetic flux due to the
appearance of parasitic bipoles at the photosphere. In any case, it is
important to mention that at the prominence body we find  vertical structures,
linked to the appearance of fingers and plumes, in an essentially horizontal
magnetic field. On the contrary, from the observational point of view it is
thought that the magnetic field is quite vertical, at least in many curtain and
hedgerow prominences formed by vertical threads.

We have showed that magnetic shear stabilizes the configuration. Performing a
quantitative study of the dependence of the instability on this parameter we
have shown that it can significantly reduce the growth-rates of the MRT unstable
modes. Thus, it has important consequences regarding the stability of the
magnetic configuration. In particular, the parameter $l/k$ in this study is
directly related to strength of the magnetic field at the prominence core and to
the change of the direction of the horizontal field with height. Strongly
sheared configurations have also a fast change of the shear angle with $z$,
which helps to stabilize the structure \citep[see][]{rudermanterradas2014}.

In our numerical experiments the transition between the corona and the
photosphere has been ignored and line-tying conditions have been applied at
$z=0$. A proper analysis of the evolution of prominences attached to the
photosphere, like the ones described above (and even suspended prominences),
should include the transition region and chromospheric layers. The physics
required to reproduce chromospheric or prominence conditions is quite complex
(partial ionization effects, radiative transport, etc..).  Moreover, thermal
effects such as conduction or radiation have been ignored in the present work
but are thought to play a relevant role on the shape of the PCTR.

It is clear that the three-dimensional configuration studied here shows very
dynamic features and the relaxation to a purely stationary magnetohydrostatic
configuration is hardly achieved, at least in the regime of parameters
considered in this work.  This is different to the two-dimensional case studied
by \citet{terradasetal2013} using essentially the same magnetic configuration.
In that case there were no MRT instabilities because the system was assumed to
be invariant with respect to the longitudinal direction (the $y-$coordinate in
our system) and modes with a component along this direction are in general the
most unstable. This indicates, once more, that interpretations based on 2D
models may miss important physics. The lack of simple magnetohydrostatic
equilibrium also agrees with the observational fact that prominences are locally
very dynamic.

Note that although we have not tried to model the prominence formation process
\citep[see][for recent advances about condensations mechanisms with magnetic
fields]{xiaetal11, xiaetal12,keppensetal2014} and the continuous supply of
material is missing, we already find a very complex dynamics. If the formation
process is investigated using levitation, injection or condensations models, it
will most likely  produce even more complex phenomena. This will make things
difficult to interpret. For this reason, we have preferred to analyze first the
evolution of an existing density enhancement in a coronal environment.

It is worth to mention that it would be interesting to study a flux rope
configuration to compare with the present shear arcade model and the
embedded direct polarity prominence in order to assess whether there are
significant differences in the results. In particular, it would be interesting
to understand the role of magnetic twist on the development of the MRT modes as
well as on the kink instability. Also, it might be interesting to perform a
detailed analysis of the modes of oscillation and to compare with observations
of oscillating prominences/filaments induced, for example, by nearby flares.
Finally, future studies should use high resolution 3D simulations to resolve the
fine structure associated to prominence threads as well as turbulent phenomena
related to nonlinearity for a better comparison with observations.

\acknowledgements J.T. acknowledges support from the Spanish Ministerio de Educaci\'on y
Ciencia through a Ram\'on y Cajal grant. R.S. acknowledges support from MINECO through a
Juan de la Cierva grant (JCI-2012-13594), from MECD through project CEF11-0012, and from
the Vicerectorat d'Investigaci\'o i Postgrau of the UIB. J.T., R.S., R.O. and J.L.B.
acknowledge the funding provided under the project AYA2011-22846 by the Spanish MICINN and
FEDER Funds.  The financial support from CAIB through the ``Grups Competitius'' scheme and
FEDER Funds is also acknowledged. M.L. gratefully acknowledges partial financial support
by the Spanish  Ministry of Economy through projects AYA2011-24808 and CSD2007-00050. 
This work contributes to the deliverables identified in FP7 European  Research Council
grant agreement 277829, ``Magnetic Connectivity through  the Solar Partially Ionized
Atmosphere" (PI: E. Khomenko). We thank the anonymous referee for his/her comments
that helped to improve the manuscript.


\end{document}